\documentclass[twocolumn]{aastex631}

\usepackage[utf8]{inputenc}
\usepackage{xcolor}
\usepackage{xspace}
\usepackage{amsmath}

\newcommand{\source}{PSR~J2147$-$3604}
\newcommand{\msun}{M$_{\sun}$\xspace}

\newcommand{\h}{^{\rm h}}
\newcommand{\m}{^{\rm m}}
\newcommand{\s}{^{\rm s}}

\newcommand{\Pobs}{$P_{\rm obs}$}

\newcommand{\Pdotobs}{$\dot{P}_{\rm obs}$}
\newcommand{\Pdotint}{$\dot{P}_{\rm int}$}
\newcommand{\TEMPOTWO}{\texttt{TEMPO2}\xspace}

\newcommand{\RNum}[1]{\uppercase\expandafter{\romannumeral #1\relax}}
\newcommand{\PRESTO}{\texttt{PRESTO}}

\begin{document}

\title{Radio and Gamma-Ray Timing of a Newly Discovered Millisecond Pulsar}

\author[0000-0003-0292-4453]{Laila Vleeschower}
\affiliation{Center for Gravitation, Cosmology, and Astrophysics, Department of Physics and Astronomy, University of Wisconsin-Milwaukee, P.O. Box 413, Milwaukee, WI 53201, USA\\}

\author[0000-0002-9409-3214]{Rahul Sengar}
\affiliation{Max Planck Institute for Gravitational Physics (Albert Einstein Institute), D-30167 Hannover, Germany \\}
\affiliation{Leibniz Universit\"{a}t Hannover, D-30167 Hannover, Germany\\}
\affiliation{Center for Gravitation, Cosmology, and Astrophysics, Department of Physics and Astronomy, University of Wisconsin-Milwaukee, P.O. Box 413, Milwaukee, WI 53201, USA\\}

\author[0000-0002-5775-8977]{Lars Nieder}
\affiliation{Max Planck Institute for Gravitational Physics (Albert Einstein Institute), D-30167 Hannover, Germany \\}
\affiliation{Leibniz Universit\"{a}t Hannover, D-30167 Hannover, Germany\\}

\author[0000-0003-4355-3572]{Colin J. Clark}
\affiliation{Max Planck Institute for Gravitational Physics (Albert Einstein Institute), D-30167 Hannover, Germany \\}
\affiliation{Leibniz Universit\"{a}t Hannover, D-30167 Hannover, Germany\\}

\author[0000-0001-6295-2881]{David L. Kaplan}
\affiliation{Center for Gravitation, Cosmology, and Astrophysics, Department of Physics and Astronomy, University of Wisconsin-Milwaukee, P.O. Box 413, Milwaukee, WI 53201, USA\\}

\author[0000-0002-8935-9882]{Akash Anumarlapudi}
\affiliation{Department of Physics and Astronomy, University of North Carolina at Chapel Hill, Chapel Hill, NC 27599-3255, USA\\}

\author[0000-0002-4678-2939]{Raghav Girgaonkar}
\affiliation{Center for Gravitation, Cosmology, and Astrophysics, Department of Physics and Astronomy, University of Wisconsin-Milwaukee, P.O. Box 413, Milwaukee, WI 53201, USA\\}

\author[0000-0002-9285-6724]{Tinn Thongmeearkom}
\affiliation{National Astronomical Research Institute of Thailand, Don Kaeo, Mae
Rim, Chiang Mai 50180, Thailand\\}
\affiliation{Jodrell Bank Centre for Astrophysics, Department of Physics and Astronomy, The University of Manchester, Manchester M13 9PL, UK\\}

\author[0000-0001-5134-3925]{G.~Y. Agazie}
\affiliation{Center for Gravitation, Cosmology, and Astrophysics, Department of Physics and Astronomy, University of Wisconsin-Milwaukee, P.O. Box 413, Milwaukee, WI 53201, USA\\}

\author[0000-0003-3294-3081]{M. Bailes}
\affiliation{Centre for Astrophysics and Supercomputing, Swinburne University of Technology, P.O. Box 218, Hawthorn, Victoria 3122, Australia\\}
\affiliation{Australian Research Council Centre of Excellence for Gravitational Wave Discovery (OzGrav)\\}

\author[0000-0001-8715-9628]{E. D. Barr}
\affiliation{Max-Planck-Institut für Radioastronomie, Auf dem Hügel 69, D-53121
Bonn, Germany\\}

\author[0000-0001-8522-4983]{R. P. Breton}
\affiliation{Jodrell Bank Centre for Astrophysics, Department of Physics and Astronomy, The University of Manchester, Manchester M13 9PL, UK\\}

\author[0000-0002-6039-692X]{H. T. Cromartie}
\affiliation{Department of Physics and Astronomy, Vanderbilt University, 2301 Vanderbilt Place, Nashville, TN 37235, USA\\}

\author[0000-0002-2185-1790]{M. E. Decesar}
\affiliation{Department of Physics and Astronomy, George Mason University, Fairfax, VA 22030, USA\\}

\author[0000-0001-8885-6388 ]{T. Dolch}
\affiliation{Department of Physics, Hillsdale College, 33 E. College Street, Hillsdale, MI 49242, USA\\}
\affiliation{Eureka Scientific, 2452 Delmer Street, Suite 100, Oakland, CA 94602-3017, USA\\}

\author[0000-0002-4175-2271]{M. Kramer}
\affiliation{Max-Planck-Institut für Radioastronomie, Auf dem Hügel 69, D-53121
Bonn, Germany\\}
\affiliation{Jodrell Bank Centre for Astrophysics, Department of Physics and Astronomy, The University of Manchester, Manchester M13 9PL, UK\\}

\author[0000-0001-5229-7430]{R. S. Lynch}
\affiliation{Green Bank Observatory, P.O. Box 2, Green Bank, WV 24944, USA\\}

\author[0000-0002-0430-6504]{E. Parent}
\affiliation{Université Grenoble Alpes, CNRS, IPAG, F-38000 Grenoble, France\\}

\author[0000-0001-5799-9714]{S. M. Ransom}
\affiliation{National Radio Astronomy Observatory, 520 Edgemont Road, Charlottesville, VA 22903, USA\\}

\author[0000-0001-9242-7041]{B. W. Stappers}
\affiliation{Jodrell Bank Centre for Astrophysics, Department of Physics and Astronomy, The University of Manchester, Manchester M13 9PL, UK\\}

\author[0000-0001-8503-6958]{J. van Leeuwen}
\affiliation{ASTRON, Netherlands Institute for Radio Astronomy, Oude Hoogeveensedijk 4, 7991 PD Dwingeloo, The Netherlands\\}
\affiliation{Leiden Observatory, Leiden University, Einsteinweg 55, 2333 CC Leiden, The Netherlands}
\affiliation{Leiden Institute for Advanced Computer Science (LIACS), Leiden University, Einsteinweg 55, 2333 CC Leiden, The Netherlands}

\begin{abstract}

We report the discovery and timing of \source, a fully recycled millisecond pulsar with a spin period of 3.2\,ms, discovered in a reprocessing of the 350-MHz Green Bank North Celestial Cap (GBNCC) survey and later serendipitously confirmed by the MeerKAT TRAPUM UHF survey of \textit{Fermi}-LAT unidentified sources. We followed up this pulsar with Murriyang/Parkes radio telescope and obtained a phase-connected timing solution spanning one year, revealing a 4.5-day binary orbit with very low eccentricity ($e \sim 1.9 \times 10^{-5}$). A follow-up search in LAT data revealed gamma-ray pulsations, allowing for a joint radio and gamma-ray timing analysis. Assuming a neutron-star mass of 1.5\,\msun, the inferred median companion mass ($\sim 0.2$\,\msun) is consistent with a helium white dwarf. The pulsar has a low dispersion measure of $17.0$~pc~cm$^{-3}$ and shows strong scintillation, leading to it being detected in $\sim 72$\% of the observations covering the 704--1174~MHz band with Murriyang. The pulsar has a narrow duty cycle of $\sim 6.5$\%, which leads to individual time of arrival (ToA) uncertainties of $\sim4.5\,\mu$s for a 10\,min integration with MeerKAT, suggesting this pulsar is well suited for inclusion in PTA experiments. 

\end{abstract}


\section{Introduction} \label{sec:intro}

Millisecond pulsars (MSPs) are neutron stars that have been spun up to rotation periods typically shorter than 30\,ms through the transfer of mass and angular momentum from a companion star  \citep{Alpar1982, Radhakrishnan+Srinivasan1982, Fabian+1983}. Since the discovery of the first MSP in 1982 \citep{Backer+1982}, more than eight hundred MSPs have been found\footnote{See \url{https://www.atnf.csiro.au/research/pulsar/psrcat/}} \citep{Manchester+2005}, the majority of which are in binary systems. The discovery and timing of binary MSPs help constrain models of the formation and evolution of compact binaries, including the nature of their companions and the mass-transfer processes responsible for recycling neutron stars \citep[e.g.,][]{Tauris+2017, Tauris+2026}.

The rotational stability of MSPs makes them targets for precision timing, enabling precise measurements of their spin, astrometric, and orbital properties (when applicable) of these systems. In favorable binaries, relativistic effects can also provide neutron-star mass measurements and stringent tests of theories of gravity \citep[e.g.,][]{Cromartie+2020, Voisin+2020, Kramer+2021}. Ensembles of the most precisely timed MSPs form pulsar timing arrays (PTAs), which are sensitive to gravitational waves at nanohertz frequencies \citep{Hellings+Downs1983, Foster+Backer1990}. Recent PTA results have provided evidence for a gravitational-wave background at these frequencies \citep[e.g.,][]{Nanograv+23, EPTA+2023, Reardon+2023, CPTA+2023, NANOGrav+2025, Miles+2025}.  

The sensitivity of PTAs depends on the number and sky distribution of MSPs, their observing cadence and timing baseline, and the precision and stability of their timing measurements \citep[e.g.,][]{Siemens+2013, Lam2018}. MSPs with narrow pulse profiles and high S/N are particularly well suited for precision timing, as the uncertainty in pulse times of arrival (ToAs) depends on the pulse width and signal-to-noise ratio (S/N). The discovery of new MSPs with narrow, stable profiles and favorable sky positions can therefore improve PTA sensitivity and increase the number of independent pulsar pairs available to measure the angular correlation induced by a gravitational-wave background.

Propagation through the turbulent interstellar medium (ISM) introduces additional challenges for precision timing, including chromatic delays and flux-density variations caused by scintillation \citep[e.g.,][]{Cordes+Shannon2010, Keith+2013, Reardon+2020}. MSPs with low dispersion measure (DM) often experience relatively weak interstellar scattering and broad diffractive scintillation bandwidths, which can cause their S/N and hence ToA precision to vary considerably between observations and across the observing band. These variations can affect both the detectability and the precision of the ToAs of a pulsar. Characterizing the scintillation properties of a newly discovered MSP is therefore important both for understanding its line of sight through the ISM and for assessing strategies for long-term precision timing.

MSPs are also prominent sources of high-energy emission. The \textit{Fermi} Large Area Telescope \citep[LAT;][]{Atwood+2009} has established a large population of gamma-ray MSPs \citep{3PC}, and radio searches of unidentified \textit{Fermi}-LAT sources have led to the discovery of many new MSPs \citep[see e.g.][]{Kerr+2012, Cromartie+2016, Wang+2021, Clark+2023, Bangale+2024, Kerr+2025, Thongmeearkom+2026}. In contrast to radio emission, gamma-ray photons are unaffected by dispersion, scattering, and scintillation in the ionized ISM. However, their low arrival rates make searches for gamma-ray pulsations dependent on accurate rotational and, for binary MSPs, orbital parameters. A phase-connected radio timing solution can therefore provide the ephemeris required to search \textit{Fermi}-LAT data for pulsations and establish a connection between the radio and gamma-ray emission.

In this work, we present the discovery of \source, a narrow duty-cycle scintillating MSP discovered through a reprocessing of archival data from the Green Bank North Celestial Cap (GBNCC) pulsar survey, a 350\,MHz all-sky survey of the Northern sky conducted from 2010-2020 \citep{Stovall+2014}. We use Parkes follow-up radio observations spanning one year to obtain a phase-connected timing solution and determine the rotational, astrometric, and binary parameters of the system. This solution enabled the detection of gamma-ray pulsations in \textit{Fermi}-LAT data, extending the timing baseline and allowing a joint radio and gamma-ray timing analysis. We discuss the implications of these measurements for the nature and evolution of the system and for future long-term timing observations of this newly identified MSP. We describe the observations and data reduction in Section~\ref{sec:data_reduction_and_searches}, the timing solution in Section~\ref{sec:timing}, and the results and their implications in Section~\ref{sec:discussion}. We draw conclusions in Section~\ref{sec:conclusion}.

\section{Observations and Data Analysis}
\label{sec:data_reduction_and_searches}

\subsection{Murriyang (Parkes) and MeerKAT}

\source\ was initially identified as a candidate in a reprocessing of the GBNCC survey \citep[for further details, see][]{Sengar+2026}, based on a 2~min observation obtained with the Green Bank Ultimate Processing Instrument (GUPPI) on 2014 December 28. The candidate was found with a spin period of $P\sim3.20$~ms, a ${\rm DM}=17.02$~pc~cm$^{-3}$, and a low S/N of 11. We found no previous record of this source in known pulsar catalogs \citep{Manchester+2005} or in the Pulsar Survey Scraper\footnote{\url{https://pulsar.cgca-hub.org}} \citep{Kaplan2022}. 

Subsequently, the MSP was serendipitously confirmed in a 10-min observation by the \textit{Fermi} Sources Working Group of TRAnsients and PUlsars with MeerKAT \citep[TRAPUM;][]{Stappers+Kramer2018} during an Ultra-High Frequency (UHF; 544--1087~MHz) survey of \textit{Fermi}-LAT sources (T. Thongmeearkom et al. in preparation, expanding on the survey described in \citealt{Thongmeearkom+2026}). The observation used 4096 frequency channels and a time resolution of 120~$\mu$s, targeting the unidentified \textit{Fermi}-LAT source 4FGL~J2146.9$-$3606 \citep{4FGLDR3,4FGLDR4}. The beamforming capabilities of the TRAPUM backend enabled by the Filterbanking BeamFormer User Supplied Equipment\citep[FBFUSE;][]{Barr2018, Chen+2021} allowed us to obtain a sub-arcsecond localization using the \textsc{SeeKAT}\footnote{\url{https://github.com/BezuidenhoutMC/SeeKAT}} package \citep{SeeKAT} following a similar procedure as in \cite{Clark+2023}. The maximum likelihood localization from SeeKAT was: right ascension (R.A.) $=21^{\rm h}47^{\rm m}01.6(1)^{\rm s}$, and declination (Dec.) $=-36^\circ04'44.6(5)''$. 

This improved position enabled follow-up observations with the Parkes/Murriyang radio telescope using the Ultra-Wideband Low \citep[UWL;][]{Hobbs+2020} receiver in conjunction with the Medusa backend. The receiver provides continuous frequency coverage from 704 to 4032\,MHz. Observations were conducted in search mode with full Stokes parameters, along with a noise diode observation before each observation, under the project code PX137. The pulsar was observed over 21 epochs between 2025 October 22 and 2026 April 01, with observation durations ranging from 39 to 124\,min (see Table~\ref{t:obs_summary}). The data were recorded with a time resolution of 64~$\mu$s, 2-bit sampling, and a frequency resolution of 500~kHz, yielding a total of 6656 frequency channels across the UWL band. Since we did not have well-constrained parameters to fold the data at that time, we searched each observation following a standard acceleration-search procedure \citep{Ransom+2002} implemented with the \textsc{PULSAR\_MINER} pipeline\footnote{\url{https://github.com/alex88ridolfi/PULSAR_MINER}} (v.1.1.5, see further details in \citealt{Ridolfi+2021}), which is based on \PRESTO\footnote{PulsaR Exploration and Search TOolkit, \url{https://github.com/scottransom/presto}}~\citep{Ransom2001,Ransom2011}. Given that the pulsar was originally discovered at 350~MHz and has a relatively low DM, we divided the UWL bandwidth into eight subbands to account for a potentially steep radio spectrum and frequency-dependent flux-density variations caused by scintillation. We searched each subband independently to maximize sensitivity to a signal that might be detectable over a limited portion of the available band. For each subband and each epoch, we adopted a deliberately conservative search interval of 16.0\,pc\,cm$^{-3}$--18.0\,pc\,cm$^{-3}$ and spin periods ranging from 3\,ms to 4\,ms. Each dedispersed time series was searched in the Fourier domain for both isolated and binary pulsar signals, using $z_{\rm max}=0$ and $z_{\rm max}=200$, respectively. Here, $z_{\rm max}=T_{\rm obs}^2a_l/(cP)$ corresponds to the maximum number of Fourier bins over which a signal is allowed to drift due to a line-of-sight acceleration $a_l$, where $T_{\rm obs}$ is the observation duration, $P$ is the pulsar period, and $c$ is the speed of light. In both searches, the powers of the first eight harmonics were summed.

\begin{figure}
\centering
\includegraphics[width=0.975\columnwidth]{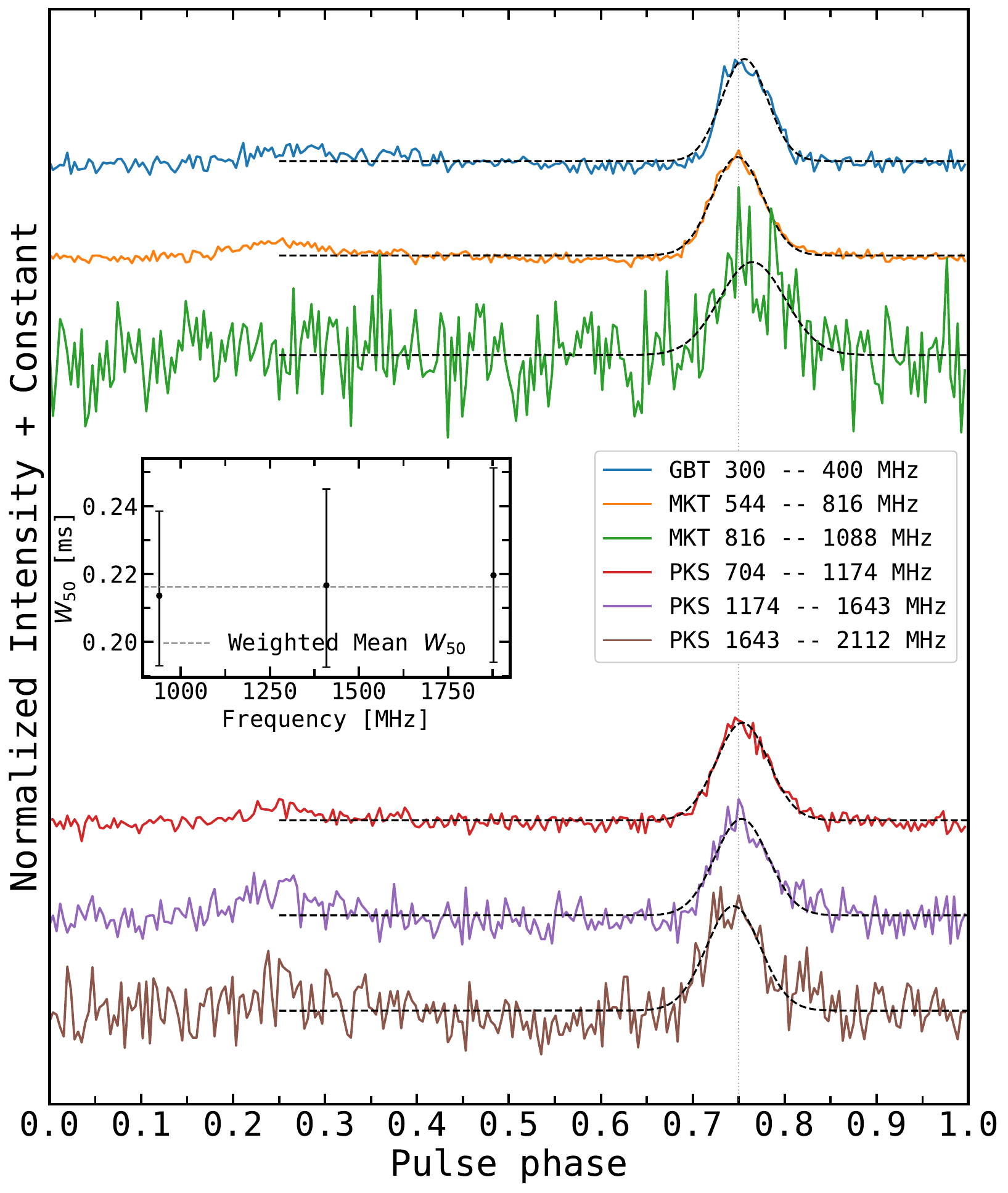}
\caption{Integrated pulse profiles of \source\ across the GBT, MeerKAT and Parkes UWL frequency bands. The profiles are normalized, aligned at the pulse peak, and vertically offset for clarity. The black dashed curves show the Gaussian fits that were used to estimate the profile widths. The inset shows the measured $W_{50}$ values for the three UWL subbands as a function of central frequency that were formed summing all observations in which the pulsar was detected in the corresponding subband; the dashed horizontal line indicates the inverse-variance weighted mean, $W_{50} = 0.22(1)$\,ms. Uncertainties were estimated from Monte Carlo noise realizations using the off-pulse rms.}
\label{f:pulse_profiles}
\end{figure}

We detected pulsed emission from \source\ between 704--1500\,MHz and mild line-of-sight acceleration in those observations, strongly suggesting the pulsar was in a binary system. We used the period-acceleration diagram method (see \citealt{Freire+2001}) to obtain starting estimates of the orbital parameters. First, we measured the barycentric observed spin period, \Pobs, and observed spin period derivative, \Pdotobs, from the observations in which we have sufficient S/N to get acceleration measured values. The \Pdotobs\ was then converted into the line-of-sight acceleration, $a_l = c(\dot{P}/P)_{\rm obs}$. This gave us first guess parameters of the orbit\footnote{Values obtained using the code \url{https://github.com/lailavc/circorbit}}: an orbital period of $P_{\rm b} \simeq 4.3$\,days, and a projected semi-major axis of only $x_{\rm p} \simeq 3.3$\,lt-s. We then used \texttt{fit\_circular\_orbit.py} from \PRESTO, returning improved orbital parameters with an orbital period of $P_{\rm b} \simeq 4.5$\,days and a projected semi-major axis of $x_{\rm p} \simeq 3.5$\,lt-s after fitting the observed spin period as a function of time, \Pobs$(t)$. 

To enable further analysis and derive a timing solution,~we obtained follow-up observations with the Parkes telescope during the 2026OCTS semester under observing program P1402 using the UWL receiver. Given the typically steep radio spectra of pulsars and the scintillation observed in the observations taken during the PX137 program, we restricted the frequency coverage to 704--2112 MHz, where the pulsar was expected to be most readily detectable. The observations had a sampling time of $64\,\mu$s, a frequency resolution of at least 125~kHz, and were coherently de-dispersed with a DM of $17.02\,$pc cm$^{-3}$. We obtained observations in both pulsar search and fold modes over $20$ epochs between $2026$ April $04$ and $2026$ September $21$, with integration times ranging from $29$ to $133\,$min (see Table~\ref{t:obs_summary}). This time the observations were split into two subbands (or three where possible) to have almost the same bandwidth per subband.

\startlongtable
\begin{deluxetable*}{ccccccccc}
\tablecaption{Summary of the radio observations used in this work. 
The dates/times are specified in UTC.\label{t:obs_summary}}
\tablehead{
\colhead{Observation} &
\colhead{Date/Time} &
\colhead{Start time} &
\colhead{$f_{\rm c}$} &
\colhead{Sampling time} &
\colhead{BW} &
\colhead{Number of} &
\colhead{$T_{\rm obs}$} &
\colhead{PSR} \\
\colhead{ID} &
\colhead{(yy-MM-dd)} &
\colhead{(MJD)} &
\colhead{(MHz)} &
\colhead{($\mu$s)} &
\colhead{(MHz)} &
\colhead{channels} &
\colhead{Length (min)} &
\colhead{Detected}
}
\startdata
\multicolumn{9}{c}{\textbf{MeerKAT}} \\
01M & 25-09-20 & 60938.02 & 816 & 120.5 & 544 & 4096  & 10 & Y \\
\multicolumn{9}{c}{\textbf{Parkes}} \\
01P & 25-10-22 & 60970.50 & 2368 & 64 & 3328 & 6656 & 39 & N \\ 
02P & 25-11-01 & 60980.42 & 2368 & 64 & 3328 & 6656 & 55 & Y \\
03P & 25-11-02 & 60981.46 & 2368 & 64 & 3328 & 6656 & 120 & Y \\
04P & 25-11-14 & 60993.50 & 2368 & 64 & 3328 & 6656 & 53 & Y \\
05P & 25-11-22 & 61001.42 & 2368 & 64 & 3328 & 6656 & 51 & N \\
06P & 25-11-29 & 61008.36 & 2368 & 64 & 3328 & 6656 & 59 & N \\
07P & 25-11-30 & 61009.32 & 2368 & 64 & 3328 & 6656 & 80 & Y \\
08P & 25-12-06 & 61015.30 & 2368 & 64 & 3328 & 6656 & 81 & N \\ 
09P & 25-12-15 & 61024.25 & 2368 & 64 & 3328 & 6656 & 84 & N \\
10P & 25-12-21 & 61030.28 & 2368 & 64 & 3328 & 6656 & 46 & Y \\
11P & 26-01-17 & 61057.26 & 2368 & 64 & 3328 & 6656 & 64 & Y \\
12P & 26-01-24 & 61064.26 & 2368 & 64 & 3328 & 6656 & 82 & Y \\
14P & 26-01-27 & 61067.25 & 2368 & 64 & 3328 & 6656 & 117 & N \\
15P & 26-01-29 & 61069.26 & 2368 & 64 & 3328 & 6656 & 124 & N \\
16P & 26-01-31 & 61071.28 & 2368 & 64 & 3328 & 6656 & 93 & Y \\
17P & 26-02-15 & 61086.19 & 2368 & 64 & 3328 & 6656 & 133 & Y \\
18P & 26-02-24 & 61095.20 & 2368 & 64 & 3328 & 6656 & 38 & N \\
19P & 26-03-01 & 61101.15 & 2368 & 64 & 3328 & 6656 & 84 & Y \\
19P & 26-03-06 & 61106.13 & 2368 & 64 & 3328 & 6656 & 90 & Y \\
20P & 26-03-20 & 61119.09 & 2368 & 64 & 3328 & 6656 & 118 & Y \\
21P & 26-04-01 & 61131.88 & 2368 & 64 & 3328 & 6656 & 100 & Y \\
22P & 26-04-04\tablenotemark{c} & 61134.83 & 2368 & 64 & 3328 & 6656 & 20 & Y \\
23P & 26-04-06\tablenotemark{c} & 61136.00 & 2368 & 64 & 3328 & 6656 & 101 & Y \\
24P & 26-04-24\tablenotemark{c} & 61154.78 & 1216 & 128 & 1024 & 8192 & 84 & Y \\
25P & 26-04-25\tablenotemark{c} & 61155.98 & 1408 & 128 & 1408 & 11264 & 92 & Y \\
26P & 26-05-02\tablenotemark{c} & 61162.00 & 1408 & 128 & 1408 & 11264 & 125 & Y \\
27P & 26-05-08\tablenotemark{c} & 61168.78 & 1408 & 128 & 1408 & 11264 & 79 & N \\
28P & 26-05-21\tablenotemark{c} & 61181.72 & 1408 & 128 & 1408 & 2816 & 79 & Y \\
29P & 26-05-28\tablenotemark{c} & 61188.67 & 1408 & 128 & 1408 & 2816 & 99 & Y \\
30P & 26-06-04\tablenotemark{c} & 61195.84 & 1408 & 128 & 1408 & 2816 & 121 & N \\
30P & 26-06-11\tablenotemark{c} & 61202.63 & 1408 & 128 & 1408 & 2816 & 83 & Y \\
31P & 26-06-17\tablenotemark{c} & 61208.76 & 1408 & 128 & 1408 & 2816 & 74 & Y \\
32P & 26-06-25\tablenotemark{c} & 61216.63 & 1408 & 128 & 1408 & 2816 & 89 & Y \\
33P & 26-07-01\tablenotemark{c} & 61222.80 & 1408 & 128 & 1408 & 2816 & 84 & Y \\
34P & 26-07-09\tablenotemark{c} & 61230.82 & 1408 & 128 & 1408 & 2816 & 87 & Y \\
35P & 26-07-16\tablenotemark{c} & 61237.55 & 1408 & 128 & 1408 & 2816 & 80 & N \\
36P & 26-07-24\tablenotemark{c} & 61245.59 & 1408 & 128 & 1408 & 2816 & 71 & Y \\
37P & 26-08-01\tablenotemark{c} & 61253.74 & 1408 & 128 & 1408 & 2816 & 133 & Y \\
38P & 26-08-06\tablenotemark{c} & 61258.73 & 1408 & 128 & 1408 & 2816 & 125 & Y \\
39P & 26-09-03\tablenotemark{c} & 61286.67 & 1408 & 128 & 1408 & 2816 & 87 & Y \\
40P & 26-09-21\tablenotemark{c} & 61304.60 & 1408 & 128 & 1408 & 2816 & 85 & Y \\
\multicolumn{9}{c}{\textbf{GBT}} \\
01G & 26-03-13 & 61112.56 & 350 & 82 & 100 & 512 & 17 & Y \\ 
02G & 26-03-14 & 61113.60 & 350 & 82 & 100 & 512 & 8 & Y \\ 
03G & 26-03-15 & 61114.62 & 350 & 82 & 100 & 512 & 7 & Y \\ 
\hline \hline 
\enddata
\tablenotetext{c}{Parkes timing campaign observations obtained under project P1402.}
\end{deluxetable*}

By directly folding the data, we detected \source\ in $\sim 72$\% of the observations at $704-1174$\,MHz with varying S/Ns. The pulsar was also detected at higher frequencies in $\sim 40$\% of the observations at $1174-1643$\,MHz and in $\sim 15$\% of the $1643-2112$\,MHz subbands, respectively. The integrated pulse profiles obtained in each frequency range are shown in Figure~\ref{f:pulse_profiles}. The combination of the pulsar's steep spectrum and strong scintillation therefore made the $704-1174$\,MHz sub-band the most sensitive and consistently detected portion of the UWL band, and we used only this frequency range for the subsequent analysis.

\subsection{Green Bank}
  
The pulsar was also observed three times with the GBT under the program GBT26A-436 as part of a campaign to confirm the pulsar candidates found in the reprocessing of the GBNCC survey \citep[see][for further details]{Sengar+2026}. These observations were carried out using the Versatile GBT Astronomical Spectrometer (VEGAS) backend centered at a frequency of 350\,MHz (300--400\,MHz bandwidth). The data set consisted of 4096 frequency channels sampled every 89~$\mu s$, with a channel bandwidth of 25~kHz, using coherent dedispersion at DM$=17.02$~pc~cm$^{-3}$. The integration times ranged from 7 to 17~min (see Table~\ref{t:obs_summary}). The pulsar was detected in all three observations.

\subsection{Optical}
\label{s:optical_obs}

\begin{figure}
\centering
\includegraphics[width=0.975\columnwidth]{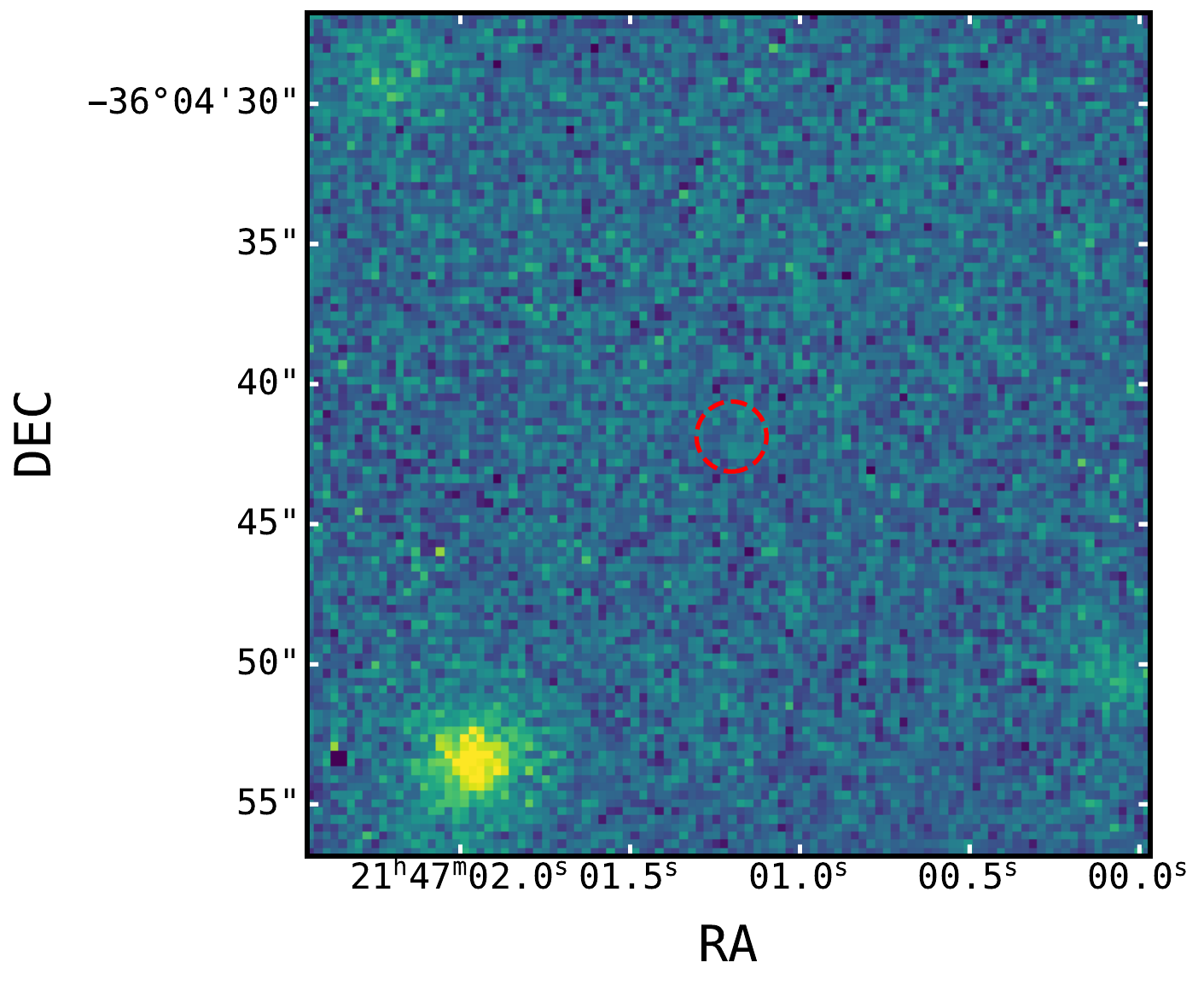}
\caption{SOAR $g$-band cutout image around the sky position of \source. The dashed red circle denotes a region of radius $1^{\prime\prime}$ around the coordinates of the pulsar.}
\label{f:soar_image}
\end{figure}

To look for the potential companion, we obtained follow-up imaging observations with the Goodman Spectrograph \citep{goodman} on the $4$-m Southern Astrophysical Research (SOAR) Telescope on 2026 June 19. We obtained five 300\,s exposures in the SDSS-\textit{g} band under excellent conditions, with seeing varying from 1\arcsec\ to 1.5\arcsec. We performed standard reduction techniques --- individual images were bias-subtracted, flat-fielded, and calibrated using \texttt{astrometry.net} \citep{astrometry}. No image subtraction was performed, given the absence of a prior counterpart. Photometry was performed using \textsc{SExtractor} \citep{sex}, and zero-point correction was done using GAIA \citep[][Data Release 3]{gaiadr3} stars as reference. We did not find a source at the location of the pulsar in any of the individual images. We stacked the individual exposures, using \textsc{Swarp} \citep{swarp}, and obtained a $5\sigma$ upper limit of $g=23.5$\,AB at the location of the pulsar. Figure \ref{f:soar_image} shows the stacked $g$-band SOAR image.

\subsection{Gamma rays}

For the gamma-ray follow-up search and timing analysis, we used data between MJD~54683 and 61296 from \textit{Fermi} LAT \citep{Atwood+2009}. We selected \texttt{SOURCE}-class photons above 100~MeV following ``Pass 8'' \texttt{P8R3\_SOURCE\_V3} instrument response functions \citep{Atwood2013+Pass8,Bruel2018+Pass8}, using the energy-dependent incidence angle, zenith angle and event-type cuts described in \citet{4FGLDR3}. Those gamma-ray photons were assigned weights, representing the probability that these photons originated from the coincident \textit{Fermi}-LAT source 4FGL~J2146.9$-$3606 instead of the diffuse background or other nearby sources \citep{Bickel2008+weights, Kerr2011+weights, Bruel2019+weights}.  These weights were computed using \texttt{gtsrcprob} from the Fermitools\footnote{Version 2.2.0, \url{https://fermi.gsfc.nasa.gov/ssc/data/analysis/documentation/}}, using the 4FGL-DR4 \citep{4FGLDR3,4FGLDR4} spectral and spatial models of the gamma-ray sky, including the \texttt{gll\_iem\_uw1216\_v13.fits} Galactic diffuse emission and \texttt{iso\_P8R3\_SOURCE\_V3\_v1.txt} isotropic emission models\footnote{\url{https://fermi.gsfc.nasa.gov/ssc/data/access/lat/BackgroundModels.html}}.

To improve computing efficiency, we introduced a cut in the photon weights, retaining photons with weight $w > 0.01$. This allows us to keep a much smaller number of high-weight photons ($\sim2,500$ photons instead of $\sim500,000$), while still retaining $99\%$ of the expected signal power \citep[which is proportional to $\sum_j w_j^2$;][]{Nieder2020+metric}.

\section{Timing Analysis and Follow up}
\label{sec:timing}

\subsection{Radio timing}
\label{sec:radiotiming}

We folded all observations in which the pulsar was detected using an initial ephemeris with \textsc{dspsr}\footnote{\url{https://dspsr.sourceforge.net}} \citep{vanStraten+2011}. The initial ephemeris included the barycentric spin period, $P$, and DM measured from the best \texttt{prepfold} detection, together with the orbital parameters obtained using \texttt{fit\_circular\_orbit.py}. We removed radio frequency interference (RFI) from the resulting archives using \textsc{clfd}\footnote{\url{https://github.com/v-morello/clfd}} \citep{Morello+2019} and processed the resulting pulse profiles using routines from the \textsc{psrchive}\footnote{\url{https://psrchive.sourceforge.net/index.shtml}} software suite \citep{Hotan+2004}.

To construct the templates used for ToA extraction, we followed a different approach for each telescope. For the MeerKAT data, we only extracted ToAs from the $544$--$816$~MHz subband, as the pulsar was only weakly detected (S/N$=10.8$) in the higher-frequency subband, and constructed the template from the resulting profile. For the GBT data, we constructed the template using the brightest observation (obs ID 03G in Table~\ref{t:obs_summary}). For the Parkes data, we coherently summed in phase the profiles from the highest-S/N observations to generate a high-S/N standard profile. For each observation, the profiles were then summed in phase over the desired number of sub-integrations and frequency channels and cross-correlated with the corresponding template to determine the ToAs.

We used the \texttt{pat} tool from the \texttt{PSRCHIVE} package to cross-correlate the selected profiles with the corresponding noise-free templates and extract topocentric ToAs. In order to maximize the number of reliable ToAs from each archive, we performed the extraction individually for each epoch, testing several levels of time averaging and selecting the configuration that provided the largest number of usable ToAs. At each trial, the pulse profiles were individually inspected to ensure a clear detection, and profiles showing significant loss of signal quality compared with the fully integrated observation were excluded. Using this procedure, multiple ToAs were obtained for the majority of the observations.

We then used the resulting ToAs to derive a preliminary timing ephemeris and repeated the entire folding and ToA-extraction procedure using this improved ephemeris. The ToAs obtained from the second iteration were used in the timing analysis presented in this work. We fitted the ToAs for the pulsar's celestial coordinates, spin parameters, and orbital parameters using \TEMPOTWO \citep{Hobbs+2006, Edwards+2006}. The timing analysis included a conversion of ToAs to the Terrestrial Time as realized from International Atomic Time, TT(TAI), and correctuons to the Solar System barycenter using the JPL~DE421 planetary ephemeris \citep{Folkner+2009}. These barycentric corrections were calculated using the astrometric parameters of the timing model. Since data from different telescopes and backends are not entirely consistent due to different delays in the signal paths of each observing system, we included two ``JUMP'' parameters for the MeerKAT and GBT ToAs, with Parkes serving as the reference dataset. We modeled the binary orbit using the ELL1 model \citep{Lange+2001}.

\begin{figure}
\centering
\includegraphics[width=0.975\columnwidth]{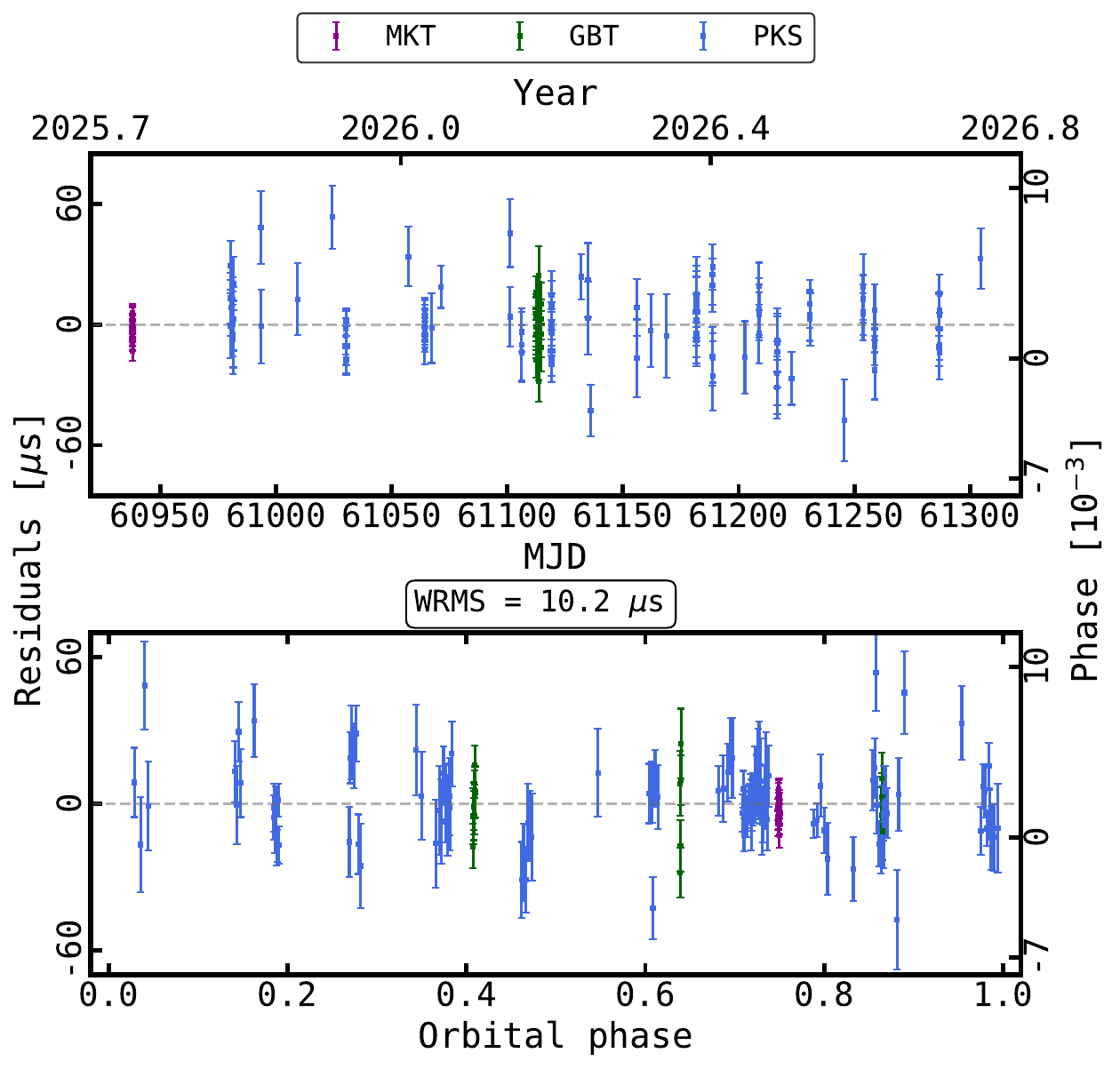}
\caption{Timing residuals of \source\ as a function of time (top) and orbital phase (bottom), where orbital phase 0 corresponds to the epoch of ascending node, $T_{\rm asc}$. The residuals were obtained using GBT, MeerKAT, and Parkes data. The blue points indicate ToAs from Parkes, while the green and purple points show TOAs from GBT and MeerKAT, respectively.}
\label{f:timing_residuals}
\end{figure}

\subsection{Gamma-ray pulsation search}

Since \source\ was found targeting a \textit{Fermi} unidentified source, 4FGL~J2146.9$-$3606, we performed follow-up searches for gamma-ray pulsations using the radio timing solution obtained in Section~\ref{sec:radiotiming}. We kept most parameters fixed and only searched for pulsations in a $2$-dimensional parameter space around the values from the radio ephemeris, $10\sigma$ in spin frequency, $f$, and $3\sigma$ in spin-frequency derivative, $\dot{f}$.

For the search, we used the $H$-test statistic \citep{DeJager1989+htest,Kerr2011+weights}.  To save computing power, we truncated the $H$ statistic computation at the $4$th out of $20$ harmonics, as pulsars typically have most power in their lower harmonics \citep{Pletsch2014+methods}.

This search resulted in a detection of gamma-ray pulsations with $H = 60.7$, without further refining the astrometric and orbital parameters. The search used $10^6$ dense trials in $f$ and $\dot{f}$. Assuming that these trials were independent results in a (highly conservative) estimated false-alarm probability of $P_{\rm FA} \approx 10^6 \times e^{-0.398405 H} = 3\times10^{-5}$ \citep{Kerr2011+weights}, corresponding to a significance of approximately $4\sigma$.

\begin{table*}[t]
\centering
\caption{Parameters for PSR~2147$-$3604 from joint timing analysis of 1.0 years of radio data and 17.8 years of gamma-ray data.  $1\sigma$ uncertainties in the final digits are reported in brackets.  Times refer to TDB. The DM-derived distance to the pulsar was estimated using the NE2025 model \citep{Ocker+Cordes2026}.}
\label{t:timing_parameters}
\begin{tabular}{ll}
\hline\hline
\multicolumn{1}{l}{Parameter} & \multicolumn{1}{c}{Value} \\
\hline
Radio data span (MJD) \dotfill & 60938 -- 61305 \\ 
Gamma-ray data span (MJD) \dotfill & 54683 -- 61296	 \\
Reference Epoch (MJD) \dotfill & 61030.3 \\ 
Number of ToAs\dotfill & 145 \\
Weighted rms Residual ($\mu s$)\dotfill & 11.2 \\
\hline
\multicolumn{2}{c}{Measured Parameters}  \\
\hline
Right Ascension (J2000), $\alpha$ (hh:mm:ss)\dotfill & 21:47:01.2087(1) \\ 
Declination (J2000), $\delta$ (dd:mm:ss)\dotfill & $-$36:04:41.824(1) \\ 
Spin Frequency, $f$ (s$^{-1}$) \dotfill & 312.63310870235(5) \\ 
1st Spin Frequency Derivative, $\dot{f}$ (Hz\,s$^{-1}$) \dotfill &  $-$5.721(2)$\times 10^{-16}$ \\ 
Dispersion Measure, DM (pc\,cm$^{-3}$)\dotfill & 17.0263(2) \\ 
Orbital Period, $P_b$ (days) \dotfill & 4.514755965(5) \\ 
Projected Semi-major Axis, $x_{\rm p}$ (lt-s) \dotfill & 3.556615(2) \\ 
Epoch of Ascending Node, $T_\textrm{asc}$ (MJD) \dotfill & 60934.6418740(5) \\ 
First Laplace-Lagrange parameter, $\epsilon_1 = e \sin \omega$ \dotfill & 1.86(14)$\times 10^{-5}$ \\ 
Second Laplace-Lagrange parameter, $\epsilon_2 = e \cos \omega$ \dotfill & $-$4.3(12)$\times 10^{-6}$ \\ 
\hline
\multicolumn{2}{c}{Derived Parameters} \\
\hline
Spin Period, $P$ (ms) \dotfill & $3.1986375472218(6)$\\
First spin period derivative, $\dot{P}$ (s s$^{-1}$) \dotfill & $5.854(2)\times10^{-21}$\\
Eccentricity, $e$ \dotfill & $1.9(1) \times 10^{-5}$ \\
Longitude of periastron, $\omega$ (deg) \dotfill & $105(2)$\\
Epoch of Periastron, $T_\textrm{0}$ (MJD) \dotfill & 60935.94(5)\\
Mass Function, $f(M_p)$ ($\rm M_{\odot}$) \dotfill & $0.002369857(4)$\\ 
Minimum companion mass, $M_{c,\text{min}}$ ($\rm M_{\odot}$) \dotfill & $0.19$ \\
Median companion mass, $M_{c,\text{med}}$ ($\rm M_{\odot}$) \dotfill & $0.22$ \\ 
$\log_{10}$(Characteristic age, yr) \dotfill & 9.85 \\
$\log_{10}$(Surface magnetic field strength, G) \dotfill & 8.19 \\
$\log_{10}$(Edot, ergs\,s$^{-1}$) \dotfill & 33.94 \\
DM-derived distance to pulsar, $d_{\rm DM}$ (kpc) \dotfill & 1.01 \\
\hline \hline
\end{tabular}
\end{table*}

\subsection{Joint radio and gamma-ray timing}

To refine the timing solution, we performed a joint timing analysis using radio and gamma-ray data. The methods used the \texttt{jrag-timing.py} script\footnote{\url{https://fermi.gsfc.nasa.gov/ssc/data/analysis/user/jrag-timing.py}} described in more detail in \citet{Burgay2024+timing}. This method uses the  \texttt{PINT}\footnote{\url{https://github.com/nanograv/PINT}} pulsar timing software to fold radio TOAs and gamma-ray photons, compute the standard test statistic for each ($\chi^2_{\rm radio}$ and $\log \mathcal{L}_{\gamma}$, respectively), and compute a joint log-likelihood function, $\log \mathcal{L}_{\rm joint} = \log \mathcal{L}_{\gamma} - 0.5 \chi^2_{\rm radio}$. 
We then explore the timing-model parameter space using this likelihood function with the affine-invariant Markov chain Monte Carlo (MCMC) sampler \texttt{emcee} \citep{Foreman-Mackey2013+emcee}. 

The best-fit joint timing solution is presented in Table~\ref{t:timing_parameters}. The timing solution from this joint analysis increases the gamma-ray test statistic to $H = 78.3$, while the radio timing residuals have a weighted rms of $10.2$~$\mu$s and a reduced $\chi^2$ of $1.7$ for $134$ degrees of freedom. The timing residuals are shown in Figure~\ref{f:timing_residuals}. The joint timing solution shows that \source\ is in a $4.5$-day, low-eccentricity orbit, with $e=2\times10^{-5}$. 

\begin{figure}
\centering
\includegraphics[width=0.975\columnwidth]{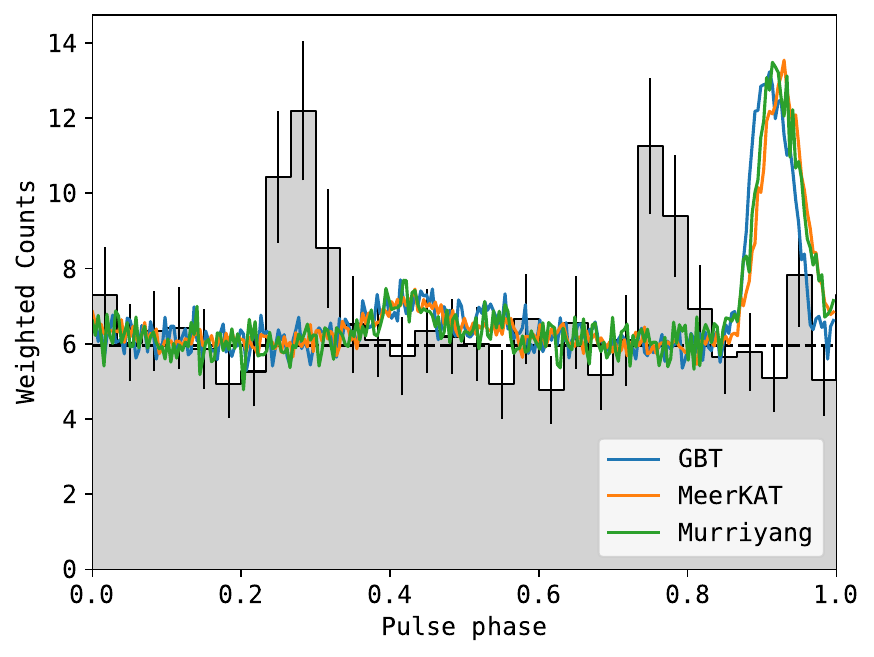}
\caption{Gamma-ray pulse profile of PSR~J2147$-$3604 and phase-aligned radio profiles. The gray histogram shows weighted counts from the \textit{Fermi}-LAT data as a function of pulse phase. The background level (dashed black line) and uncertainties are estimated from the photon weight distribution as described in \citet{3PC}. The colored lines show the radio pulse profiles, with normalization and baseline chosen (arbitrarily) to match the gamma-ray pulse.}
\label{f:fermi_profile}
\end{figure}

The resulting gamma-ray pulse profile is shown in Figure~\ref{f:fermi_profile}, alongside the radio profiles. We find a two-peaked gamma-ray profile, with a peak separation that is similar to that seen between the two radio components. It is unclear whether the gamma-ray peaks lead or lag behind the radio components, with phase offsets of 0.15 or 0.35 rotations, respectively. The former case, with gamma-ray emission leading radio pulses, has only been seen in recycled pulsars \citep{3PC}. Phase aligned radio and gamma-ray emission has been seen in several pulsars \citep{KramerJohnston2026}, interpreted as being due to radio emission from beyond the light cylinder, in the equatorial current sheet where the gamma rays are thought to be emitted. We see no significant radio emission at the phases of the gamma-ray pulses here.

\subsection{Constraints on the binary companion}

\begin{figure}
\centering
\includegraphics[width=0.975\columnwidth]{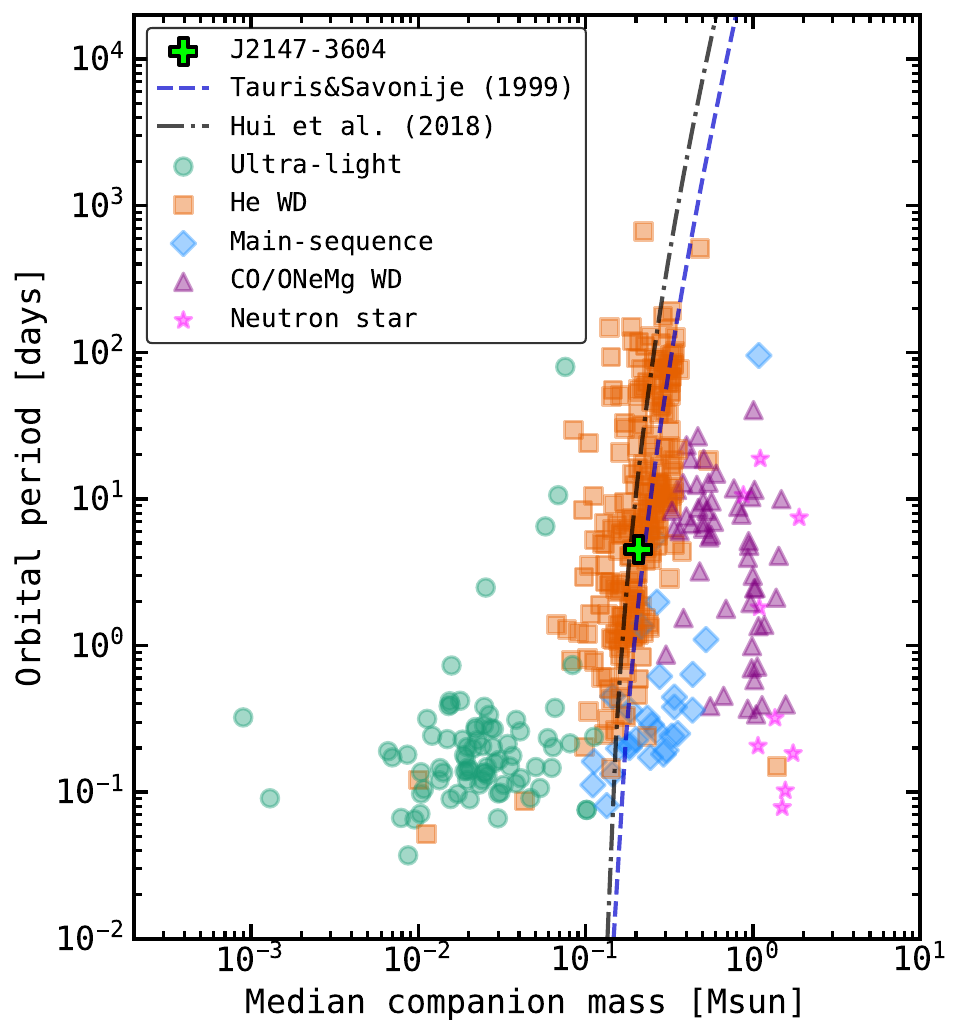}
\caption{Orbital period as a function of median companion mass for Galactic binary MSPs, shown according to their companion type. The pulsar properties were obtained from the ATNF Pulsar Catalogue. \source\ is highlighted in a lime cross. The blue dash-dotted curve shows the theoretical orbital period–white-dwarf mass relation predicted by \citet{Tauris+1999} for MSP–He WD binaries resulting from low-mass X-ray binary evolution, while the black dash-dotted curve represents the relation for MSP–He WD systems by \citet{Hui+2018}.}
\label{f:pbvsmass_msps}
\end{figure}

From the mass function derived from the timing solution, we obtain a minimum companion mass of $M_{c,\text{min}}=0.19$~\msun and a median companion mass of $M_{c,\text{med}}=0.22$~\msun, assuming a fully recycled neutron-star mass of 1.5~\msun \citep{Ozel+Freire2016}. For an orbital inclination of $i=30^{\circ}$, the corresponding companion mass is approximately 0.41~\msun. Based on the orbital period--white-dwarf mass relation of \citet{Tauris+1999} \citep[see also][]{Hui+2018}, the companion is most likely a helium-core white dwarf (He WD), with an expected mass of $\sim0.22$~\msun. Given the assumed pulsar mass this implies an orbital inclination of $i\simeq60^\circ$ (see Figure~\ref{f:pbvsmass_msps}).

The non-detection of an optical counterpart provides an additional constraint on the properties of the companion. Using the $5\sigma$ limiting magnitude of $g \ge 23.5$ obtained from the SOAR observations (Section~\ref{s:optical_obs}), we placed constraints on the effective temperature of the companion using white-dwarf cooling models\footnote{\url{http://www.astro.umontreal.ca/~bergeron/CoolingModels}} \citep{1995PASP..107.1047B,2006AJ....132.1221H}. 

We estimated the extinction toward \source\ using two independent methods. First, we used the two-dimensional SFD dust model \citep{1998ApJ...500..525S}, through the \texttt{dustmaps} Python package \citep{Green2018}, to get a line-of-sight extinction of $A_V=0.09$ and a corresponding $g$-band extinction of $A_g=0.07$. Combining the extinction-corrected limiting magnitude with the DM-based distance to \source\ (see~Table\ref{t:timing_parameters}) yields an absolute-magnitude limit of $M_g=13.5$. For a companion mass of $0.2$~\msun, the cooling models imply $T_{\rm eff}\lesssim5760$~K, while for a companion mass of $0.4$~\msun, representative of a lower orbital inclination, we obtain $T_{\rm eff}\lesssim6000$~K.

As an independent estimate, we estimated the line-of-sight hydrogen column density, $N_{\rm H}$, from the pulsar DM using the empirical relation of \citet{He_2013}. For ${\rm DM}=17.02$~pc~cm$^{-3}$, we obtain $N_{\rm H}=5.11\times10^{20}$~cm$^{-2}$. Using the $N_{\rm H}$--extinction relation of \citet{1995A&A...293..889P} gives $A_V=0.3$ and $A_g=0.24$, corresponding to an absolute-magnitude limit of $M_g=13.3$. These values differ slightly from those estimated from the dust model, however, both methods give similar effective temperature estimates. From these values, for a median companion mass of $0.2\,M_{\odot}$ we get $T_{\rm eff} \leq 5900\,$K and for a companion mass of $0.4\,M_{\odot}$, we get $T_{\rm eff} \leq 6800\,$K.

\subsection{First spin period derivative}

By extending the timing baseline through the inclusion of gamma-ray data, the joint radio and gamma-ray timing solution provides a precise measurement of the period $P\approx3.2$~ms and period derivative, $\dot{P}\approx5.8\times10^{-21}$~s~s$^{-1}$. This value is consistent with the low spin-down rates typically observed for recycled MSPs.

The observed period derivative (\Pdotobs) can be affected by acceleration effects in addition to the intrinsic spin-down (\Pdotint). These contributions can be expressed as \citep{Damour+Taylor1991}:

\begin{equation}
\label{eq:Pdot_obs}
\left(\dfrac{\dot{P}}{P}\right)_{\rm obs} = \left(\dfrac{\dot{P}}{P}\right)_{\rm int} + \dfrac{a_{\rm G} + a_{\rm PM}}{c}.
\end{equation}

\noindent Here, $a_{\rm G}$ is the difference between the accelerations of the Solar System and the pulsar in the field of the Galaxy, and $a_{\rm PM} = \mu^2 D$ is the centrifugal acceleration arising from the transverse Doppler effect, associated with the proper motion of the pulsar $\mu$, and its distance from Earth $D$ \citep{Shklovskii1970}.

We calculate the acceleration imparted by the Galaxy to \source, using equations 16 and 17 in \citet{Lazaridis+2009}, its Galactic coordinates $l=8.1743^\circ$, $b=-50.1074^\circ$, the values of the Solar motion in the Galaxy $R_0=8.275$~kpc, and $\Theta_0=240.5$~km~s$^{-1}$ \citep{Guo+2021,Grav+2021}. At the DM-derived distance to the pulsar of 1.01~kpc (see Table~\ref{t:timing_parameters}), the pulsar lies at a Galactic height of $|z|\simeq0.77$~kpc. We then estimate

\begin{equation}
\dot{P}_{\rm Gal}\simeq-3.1\times10^{-22}\ {\rm s~s^{-1}}.
\end{equation}

\noindent The magnitude of the Galactic contribution is therefore only $\sim5$\% of \Pdotobs.

The Shklovskii contribution cannot be determined because the current timing data do not provide a significant measurement of the pulsar proper motion. This correction may nevertheless be important given the small \Pdotobs. In the absence of a proper-motion measurement, \Pdotobs\ itself can be used to place an upper limit on the Shklovskii contribution. Requiring a non-negative intrinsic period derivative, $\dot{P}_{\rm int}\geq0$, gives

\begin{equation}
\dot{P}_{\rm Shk} \leq \dot{P}_{\rm obs}-\dot{P}_{\rm Gal}
\simeq6.2\times10^{-21}~{\rm s~s^{-1}}.
\end{equation}

\noindent At the adopted NE2025 distance, and using

\begin{equation}
\dot{P}_{\rm Shk}=P\frac{\mu^2D}{c},
\end{equation}

\noindent this corresponds to an upper limit on the total proper motion of

\begin{equation}
\mu\lesssim28~{\rm mas~yr^{-1}},
\end{equation}

\noindent or, equivalently, a transverse velocity of $v_\perp\lesssim134$~km~s$^{-1}$. This upper limit is larger than, but consistent with the transverse velocities measured for field MSPs.  \citet{Shamohammadi+2024} measured a mean transverse velocity of 78(8)~km~s$^{-1}$ for 77 MSPs using $\sim 3$~yr of MeerKAT observations. For a transverse velocity of $v_\perp = 78$~km~s$^{-1}$, the corresponding Shklovskii contribution would be $\sim 36$\% of the \Pdotobs. Including proper motion in our joint radio/gamma-ray timing analysis resulted in uncertainties that are similar in magnitude to this upper limit. A longer timing baseline will therefore be required to measure the proper motion and obtain a robust determination of \Pdotint. If the observed value is used without applying a Shklovskii correction, the standard magnetic-dipole relations give a characteristic age of $\tau_{\rm c}\simeq7$~Gyr, a surface dipolar magnetic-field strength of $B_{\rm s}\simeq1.5\times10^8$~G, and a spin-down luminosity of $\dot{E}\simeq8.7\times10^{33}$~erg~s$^{-1}$, assuming a neutron-star moment of inertia of $I=10^{45}$~g~cm$^2$. These values should therefore be regarded as apparent, rather than the intrinsic spin-down parameters until the Shklovskii contribution can be constrained. However, the inferred characteristic age and magnetic-field strength are characteristic of recycled MSPs.

The gamma-ray energy flux above 100~MeV from 4FGL DR4 \citep{4FGLDR4} of $G = (6.3 \pm 1.6)\times10^{-13}$~erg~cm$^{-2}$~s$^{-1}$ corresponds to a luminosity of $L_{\gamma} = (7.7 \pm 2.0)\times10^{31}$~erg~s$^{-1}$, assuming the NE2025 distance. This represents an efficiency of around $L_{\gamma} / \dot{E} = (0.9\pm0.2)\%$ of the apparent spin-down power, which is small but not unprecedented for a gamma-ray MSP \citep{3PC}, and accounting for a Shklovskii contribution would increase this value by reducing $\dot E$. 

\begin{figure}
\centering
\includegraphics[width=1\columnwidth]{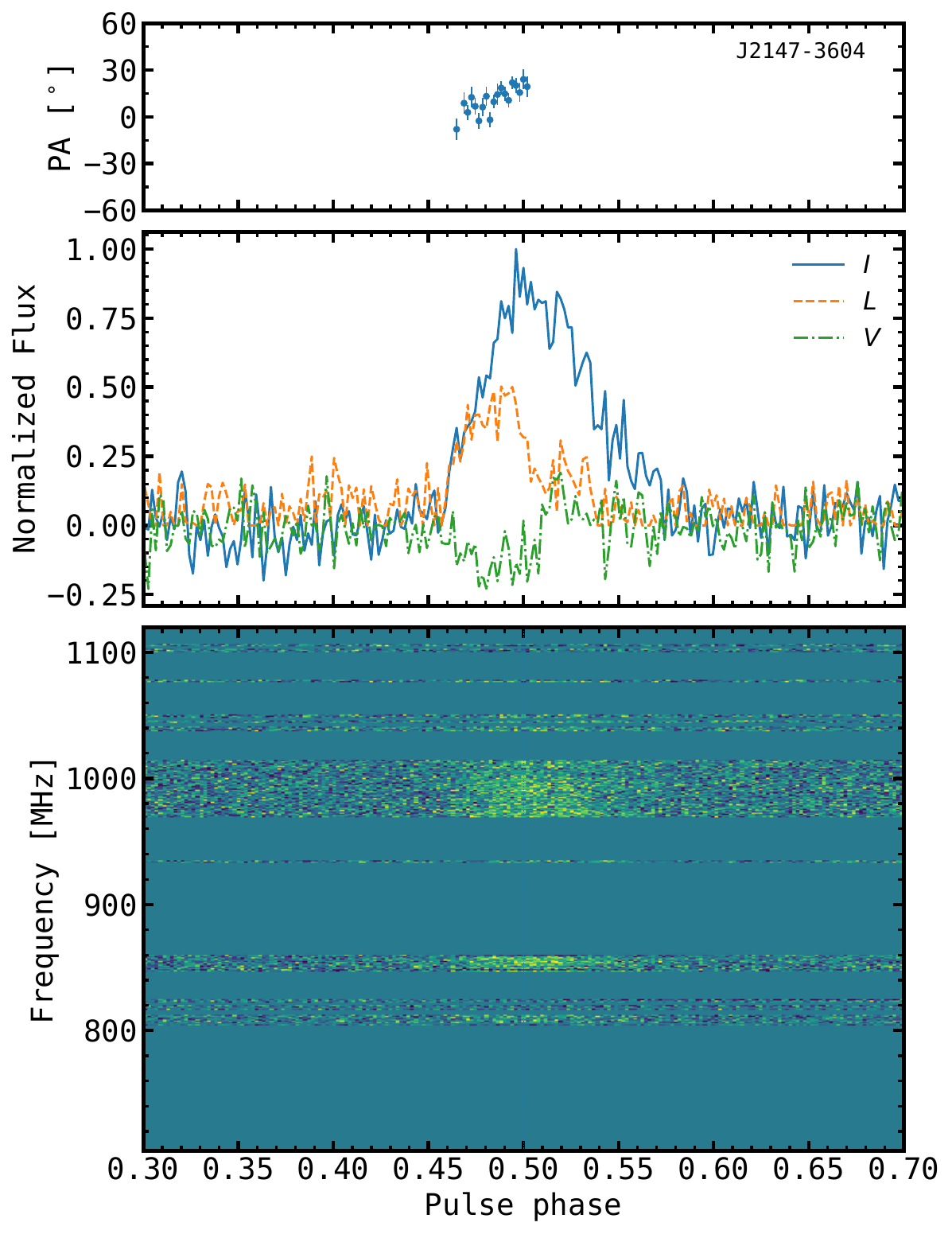}
\caption{Top panel: Position angle (PA) across the leading edge of the pulse profile, derived from the linearly polarized emission. Middle panel: Total intensity (Stokes $I$; blue), linear polarization ($L$; orange), and circular polarization (Stokes $V$; green) as a function of pulse phase. Bottom panel: Frequency-resolved pulse intensity as a function of pulse phase, illustrating the evolution of the pulse structure across the 704--1120~MHz observing band.
Frequency channels affected by persistent RFI have been masked.}
\label{f:polarization}
\end{figure}

\subsection{Polarization}

We performed flux-density and polarization calibration of the Parkes observations of \source\ following standard procedures. Each pulsar observation was accompanied by a $\sim 2-$min observation of a pulsed noise diode, which was used to calibrate the differential gain and phase between the polarization channels. The noise-diode observations were identified as polarization calibrators (\texttt{PolnCal}) and were cleaned of RFI using \texttt{clfd}, followed by manual excision of remaining contaminated frequency channels and sub-integrations using \texttt{PSRCHIVE}'s suite \texttt{pazi}.

Absolute flux-density and polarimetric calibration were performed using the \texttt{pac} routine in \texttt{PSRCHIVE}. We constructed a calibration database containing the noise-diode observations, archival flux-calibrator observations of PKS~B1934$-$638, which is regularly observed at Parkes as part of the Parkes Pulsar Timing Array project \citep[PPTA;][]{Manchester+2013}, and the appropriate measurement-equation (\texttt{pcm}) solution for the UWL receiver. The latter accounts for the frequency-dependent instrumental polarization response. The full calibration solution was applied to the pulsar observations before restricting the calibrated data to 704--1120~MHz, where \source\ is detected with the highest S/N.

We investigated the Faraday rotation of the linearly polarized emission by measuring the rotation measure (RM) using the \texttt{rmfit} routine of \texttt{PSRCHIVE}. Individual-epoch RM searches were initially performed by maximizing the linearly polarized intensity after correcting the Stokes parameters for trial values of RM. We additionally used the differential position-angle refinement implemented by \texttt{rmfit}, which estimates the residual position-angle offset between two frequency sub-bands and iteratively refines the RM. Most individual observations did not provide robust RM constraints because of the limited S/N. The observation from 2026 January 24 provides the strongest individual RM constraint. We obtain an observed RM of $18.5\pm5.1~{\rm rad~m^{-2}}$. We also estimated the ionospheric contribution using \texttt{FRion}\footnote{\url{https://frion.readthedocs.io/en/latest/}} \citep{Van+Cameron2025}, which uses \texttt{RMextract} to calculate the time-dependent ionospheric Faraday rotation, obtaining a mean ionospheric RM of $-2.57$~\rm rad\,m$^{-2}$ during the observation. After subtracting this contribution, we obtain an ionosphere-corrected RM of approximately $21.1~{\rm rad~m^{-2}}$, consistent with the range of RM values observed for low-DM pulsars \citep[see Figure~4 of][]{Wang+2024}. The resulting polarization profile and frequency--phase diagram are shown in Figure~\ref{f:polarization}. The integrated profile exhibits a linear polarization fraction of $\sim 42$\% and an absolute circular polarization fraction of $\sim 18$\%. 

We also examined the co-added data from the three observations with the highest S/N and polarization information: 2025 December 21, 2026 January 24, and 2026 March 20, obtaining RM estimates of $18.2$ and $15.6~{\rm rad\,m^{-2}}$ from the restricted RM search and differential position-angle refinement, respectively. Despite the higher S/N, the method-dependent results and uncorrected epoch-to-epoch ionospheric contributions prevented a more robust RM determination. We therefore adopted the 2026 January 24 observation for our RM measurement.

\section{Discussion}
\label{sec:discussion}

\subsection{Scintillation}

\begin{figure}
\centering
\includegraphics[width=0.975\columnwidth]{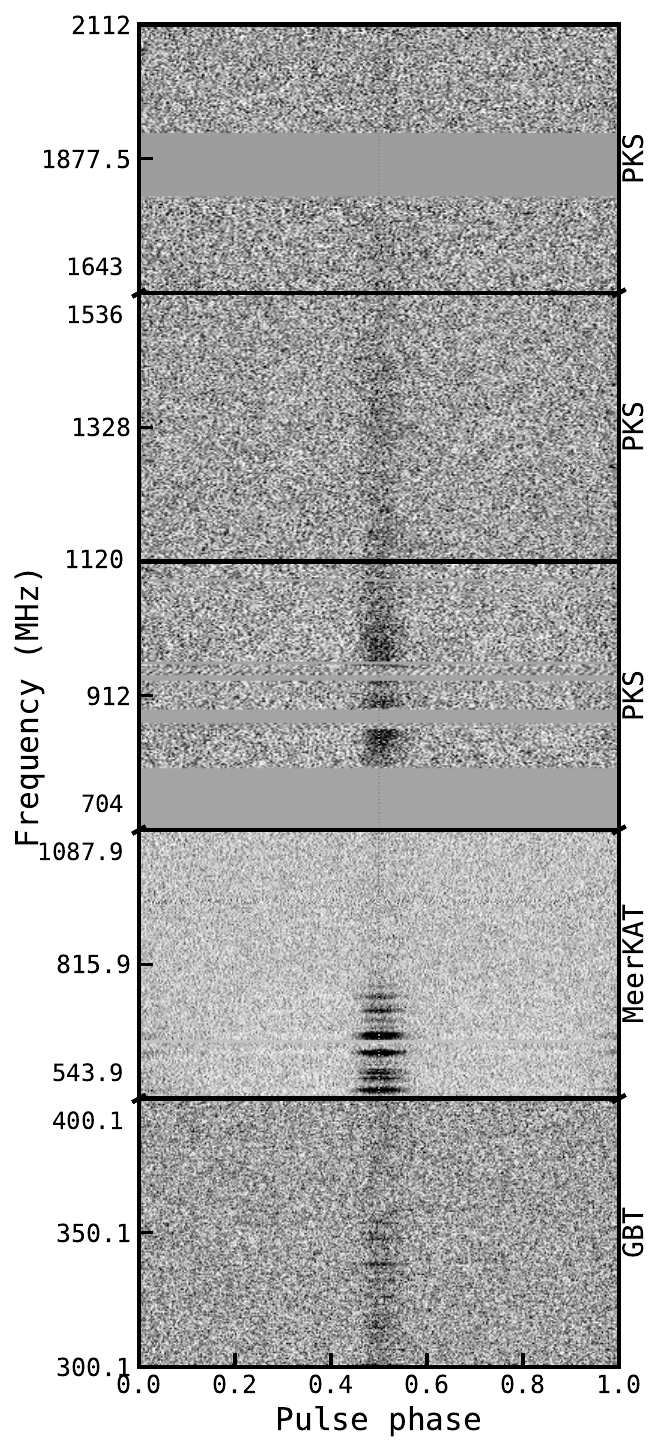}
\caption{Frequency–pulse-phase plots of \source\ obtained from observations with GBT, MeerKAT, and Parkes. The GBT panel was produced by combining the data from the three observations, while the MeerKAT panel shows the full UHF band from a single observation. The Parkes panels show the three brightest epochs in the corresponding subbands (observation IDs 12, 23, and 37, respectively). The observations are ordered by increasing observing frequency from bottom to top. Diagonal marks indicate discontinuities in frequency coverage between the different observing telescopes/bands. Frequency channels affected by RFI have been masked.}
\label{f:freqvsphase}
\end{figure}

As the signals from pulsars traverse the ISM, they experience phase modulations due to its turbulent and inhomogeneous nature. The intensity of the pulsar signal fluctuates on a variety of bandwidths and timescales. This effect is known as interstellar scintillation. 

Given the low DM of \source, relatively weak interstellar scattering and broad diffractive scintillation bandwidths are expected. Using the NE2025 electron density model, we estimate a scintillation bandwidth and timescale of $\sim 6.7$\,MHz and $\sim 661$\,s at 1\,GHz, respectively. Scaling these values to the central frequency of the lower half of the MeerKAT UHF band, $\sim 680$\,MHz, yields an expected scintillation bandwidth $\Delta \nu_{d} \sim 1.4$\,MHz and a scintillation timescale $\Delta t_d \sim 416$\,s. We have used the sensitive MeerKAT observation (see Table~\ref{t:obs_summary}) to measure the scintillation properties of the pulsar via the auto-correlation function (ACF) of the dynamic spectrum (DS). The DS was formed from the on-pulse emission after subtracting the off-pulse baseline and masking the RFI in the affected channels and subintegrations. The 2D ACF was calculated using a Fourier-domain correlation method, and Gaussian functions were fitted to the frequency and time slices of the ACF to estimate $\Delta \nu_{d}$ and $\Delta t_d$. 

We obtain $\Delta \nu_{d} = 4.5(3)$\,MHz and $\Delta t_d = 925(21)$~s. Since the inferred $\Delta t_d$ is larger than the total observation length, the scintillation timescale cannot be reliably constrained, and we conservatively take $\Delta t_d \gtrsim 10$\,min. The usable bandwidth of $\sim 271$\,MHz spans $\sim60$ scintillation bandwidths ($B/\Delta\nu_d\simeq60$). The effective number of scintles sampled during the observation can then be estimated as

\begin{equation}
    N_{\rm scint}=
\left(1+\eta\frac{T_{\rm obs}}{\Delta t_d}\right)
\left(1+\eta\frac{B}{\Delta\nu_d}\right),
\end{equation}

\noindent where $\eta$ is the scintle filling factor. Using the ACF-derived value of $\Delta t_d$ for this estimate and assuming $\eta = 0.1$--$0.3$ \citep[e.g.,][]{Cordes1986, Turner+2024}, we obtain $N_{\rm scint} \approx 7.4$--$22.4$. For statistically independent scintles, the expected modulation index due to diffractive scintillation (DISS) scales as $m_{\rm DISS}\simeq N_{\rm scint}^{-1/2}$ \citep[e.g.,][]{Cordes+Lazio1991,Jankowski+2018}, giving $m_{\rm DISS}\sim0.2$--$0.4$ Since $\Delta t_d$ is not resolved by the observation, its contribution to $N_{\rm scint}$ remains uncertain, and these values should be considered approximate. 

For a $\mathrm{DM}=17.02~{\rm pc~cm^{-3}}$, the pulsar falls within the $15$--$25~{\rm pc~cm^{-3}}$ DM bin shown in Figure~6 of \citet{Gitika+2023}, where the reported modulation indices are typically $m \sim 1.1$--$1.3$ over the frequency range $944$--$1625$\,MHz, albeit with substantial scatter. Our estimated modulation index at $\sim680$\,MHz is lower than the central values in this distribution, likely due in part to averaging over the large observing bandwith relative to the scintillation bandwidth. The usable bandwidth of $\sim271$\,MHz spans approximately $60$ scintillation bandwidths for $\Delta\nu_d \approx 4.6$\,MHz, leading to significant averaging over independent scintles. Scaling the scintillation parameters as $\Delta\nu_d \propto \nu^{4\text{--}4.4}$ and $\Delta t_d \propto \nu^{1.2}$ yields an estimated modulation index of $m \sim 0.4$--$0.6$ at 944\,MHz, which is broadly consistent with the range shown in Figure~6 of that paper once the observed scatter is taken into account.

\begin{figure}
\centering
\includegraphics[width=1\columnwidth]{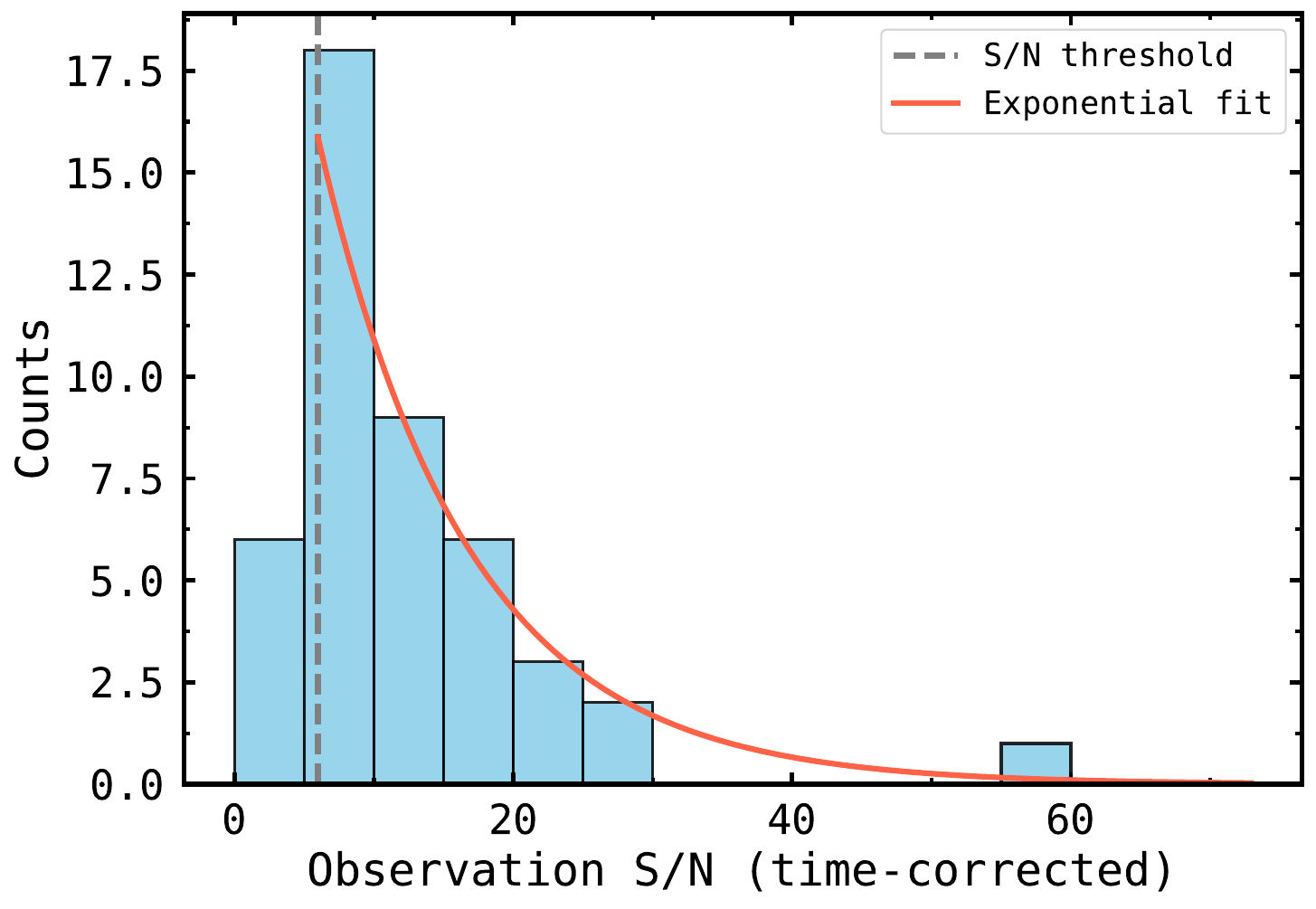}
\caption{Histogram of the S/N of all Parkes observations for PSR~J2147--3604. The overlaying curve is an exponential function that has been fit to the data. The grey dotted line at raw S/N of 7 represents our threshold of detection.}
\label{f:observation_snr}
\end{figure}

We also used the Parkes observations in the frequency range 704--1174~MHz to investigate the scintillation properties of this pulsar. First, we tried a similar analysis to that performed for the MeerKAT observation using the observation with the highest S/N (time-corrected S/N $\sim 60$, observation from 2026 January 24), but the DS S/N was only $\sim 1.5$ per time--frequency bin,  estimated from the mean on-pulse excess relative to the variations in the off-pulse intensity. This was insufficient for a reliable measurement of the scintillation parameters from the DS and its ACF. Instead, we looked at the variations in the observation S/N to find evidence of scintillation. Figure~\ref{f:observation_snr} shows a histogram of the total observation S/N of the Parkes observations in the frequency range 704--1174\,MHz. The S/N was normalized in terms of the duration of the observation. The histogram shows an approximately exponential tail, consistent with the intensity variations expected from diffractive scintillation.

The modulation index was computed as

\begin{equation}
    m = \frac{\sigma_{S}}{\langle S \rangle},
\end{equation}

\noindent where $\sigma_{S}$ is the standard deviation of the corrected S/N values, and $\langle S \rangle$ is their mean. We obtain $m \approx 1$, indicative of strong diffractive scintillation.

This result complements the scintillation analysis of the MeerKAT UHF observation. In the MeerKAT data, the large observing bandwidth relative to the scintillation bandwidth and the limited observation duration imply that many scintles are averaged over in frequency while the scintillation pattern is not fully resolved in time which leads to a reduced modulation index. In contrast, the larger modulation index observed in the Parkes data suggests that fewer independent scintles are being averaged over within a single observation, resulting in stronger apparent intensity variations between epochs.

\subsection{Non-detection of the pulsar in previous surveys}

\source\ is located at high Galactic latitude ($b=-50.1^\circ$), a region of the sky that remains relatively underexplored compared with the Galactic plane, where many of the deepest pulsar surveys conducted with major radio facilities have preferentially targeted, since the pulsar density is expected to be highest in this region. In particular, most surveys covering the position of \source\ have been conducted at 1--2~GHz \citep[e.g.,][]{Manchester+2001,Keith+2010,Padmanabh+2023}.

In our 1174--1643~MHz Parkes observations, \source\ is found in only $\sim 44\%$ of epochs, with an average S/N of 13.5 (ranging from 7.2 to 35.0) for detections and an average integration time of $\sim 97$~min. The most relevant previous L-band survey covering this region of the sky is the high-Galactic-latitude component of the High Time Resolution Universe survey (HTRU-S HighLat; \citealt{Keith+2010}). HTRU-S HighLat observed the southern sky using the Parkes multibeam receiver at a central frequency of 1352~MHz, with an integration time of only 270~s per pointing and a nominal limiting flux density of HTRU-S HighLat is $S_{\rm min}\simeq0.22$~mJy. For comparison, our Parkes observations yielded an average flux density of $S_{1352} = 0.06$~mJy which is a factor of 3.3 below the nominal survey detection threshold, the brightest Parkes observation yielded only $S_{1352}\simeq0.18$~mJy. Furthermore, the nearest HTRU-S HighLat pointing was offset by $\sim6'$ from the pulsar's position, reducing the telescope gain to $\sim60\%$ of its on-axis value and increasing the effective detection threshold to $\sim0.33$~mJy, assuming a Gaussian beam response. The previous non-detection is therefore consistent with the survey sensitivity and reduced gain at the pulsar's position, with scintillation contributing to its variable detectability.

\source\ was only found during the reprocessing of the GBNCC survey using an optimized FFT-based periodicity search. This approach has recovered a number of pulsars missed by the original processing, including faint sources and those with narrow pulse profiles \citep[see e.g.,][]{Sengar+2023, Sengar+2025, Sengar+2026}. The discovery of this MSP therefore demonstrates the importance of reprocessing archival surveys with improved search techniques, candidate sifting, folding, and classification methods. 

\subsection{Timing Properties}

The small duty cycle of \source\ of only 6.5\% at $\sim 940$~MHz places this pulsar as a good candidate for high-precision timing. The ToA precision that can be achieved can be estimated from the current observations, although a longer radio timing baseline will be required to determine its stability over time.

According to the NE2025 model, the predicted scattering timescale, responsible for temporal broadening of the pulse profile at 1~GHz, is $\tau_{\rm scat,1\,GHz} = 0.0276~\mu{\rm s}$, indicating that the pulse profile width is unaffected by scattering. This value is consistent with the scattering timescales of MSPs found at high-Galactic latitudes due to their low DM values. From the MeerKAT 10~min observation at 670~MHz, we obtained a mean ToA uncertainty of $\sim 4.5$~$\mu$s when divided into 20 subintegrations2. Assuming the uncertainties scale as $T_{\rm obs}^{-1/2}$ and dividing the observation into eight ToAs would increase the integration time per ToA from $\sim30$ to $\sim75$~s and reduce the expected mean uncertainty to $\sim2.8~\mu$s. Scaling to a 2048-s observation, with the eight ToAs per observation, gives a mean ToA uncertainty of $\sim1.6~\mu$s. These values are comparable to the few-$\mu$s sub-banded ToA uncertainties measured for pulsars in the MeerKAT Pulsar Timing Array \citep[MPTA; ][]{Miles+2025}, suggesting that \source\ may be capable of similar timing precision during favorable scintillation states. 

The timing precision of \source\ is strongly dependent on its observed flux density. The pulsar is not detected in $\sim 28\%$ of the Parkes 704--1174~MHz observations, while in other epochs only a single ToA per observing epoch can be measured, whereas as many as 20 ToAs were obtained from the MeerKAT observation. This variability is therefore likely to produce substantial epoch-to-epoch differences in both the number and precision of the available ToAs. Longer observations can partly compensate for this effect. The MPTA, for example, uses integration times between $256$ and $2048$~s, adjusted according to the sensitivity required to reach a ToA uncertainty of approximately $1\,\mu$s \citep{Miles+2025}.

Although the radio timing observations of \source\ currently span one year, the resulting phase-connected radio ephemeris enabled us to obtain a join radio and gamma-ray timing analysis using 17.8~yr of \textit{Fermi}-LAT data. The substantially longer timing baseline allows us to better constrain parameters that remain covariant over the short radio baseline, particularly the spin-frequency derivative and the pulsar position. In the joint solution, we measure a precise spin-frequency derivative of $\dot{f}=-5.721(2)\times10^{-16}$~Hz~s$^{-1}$, while the positional uncertainties are reduced by factors of $\sim2$ and $\sim8$ in right ascension and declination, respectively, compared with the radio-only solution. These improved constraints reduce uncertainties in the timing model that would otherwise require several years of radio observations, although a longer radio baseline is still needed to determine the stability of the timing residuals and characterize any chromatic or intrinsic noise.

Given the timing precision expected from future MeerKAT observations, we also investigated the timescale over which continued radio timing could constrain relativistic effects in the binary. We simulated ToAs over different timing baselines using \textsc{pint} \citep{Luo+2018} and the currently estimated orbital parameters, focusing on the relativistic periastron advance ($\dot{\omega}$), measured through the time derivatives of the first and second Laplace-Lagrange parameters ($\dot{\epsilon}_1$ and $\dot{\epsilon}_2$) in the ELL1 binary model, and the orbital-period derivative predicted by general relativity ($\dot{P}_B$). We assumed a cadence similar to that generally used in the MPTA, consisting of a 2048-s observation every two weeks, divided into eight ToAs with an uncertainty of $1.6~\mu$s per ToA. Under these assumptions, we find that a significant ($>3\sigma$) measurement of the relativistic periastron advance is expected to require a timing baseline of $\gtrsim27$~yr. The orbital-period derivative remains below this significance over the 30-yr interval explored in our simulations. A detection of the Shapiro delay is more difficult to predict, as its amplitude depends strongly on the orbital inclination, which is currently unconstrained for \source. An interesting comparison is PSR~J0740+6620, an MSP in a binary with similar orbital properties, with $P_B\simeq4.77$~d, $e\simeq5\times10^{-6}$, and a $\sim0.2\,M_\odot$ white-dwarf companion. A strong Shapiro-delay signature can be measured in PSR~J0740+6620 because the system is viewed nearly edge-on ($i\simeq87^\circ$), allowing precise constraints on both component masses \citep{Cromartie+2020, Fonseca+2021}. If \source\ has a sufficiently high inclination, continued high-precision timing may allow the Shapiro delay, and hence the component masses, to be constrained.

The position of \source\ places it in a close angular separation to PSR~J2124--3358, the nearest pulsar in the NANOGrav 15-yr dataset \citep{NANOGrav+2025}. The two sources are separated by $\sim 5^{\circ}$. Consequently, the inclusion of \source\ in the NANOGrav PTA would introduce a closely separated pulsar pair that samples the small-angular-separation region of the Hellings-Downs correlation. At this separation, the expected Hellings–Downs correlation coefficient is $\sim 0.47$, near the largest positive correlation predicted by the Hellings–Downs curve for a pair of different pulsars \citep{Hellings+Downs1983}. \source\ may therefore provide useful additional sampling of the strongly correlated, small-separation portion of the Hellings–Downs curve, although its ultimate contribution to gravitational-wave sensitivity will depend on its achievable timing precision, observing cadence, timing baseline, and intrinsic noise properties. On the other hand, \source{} is not in close angular proximity to any of the individual supermassive black hole binary (SMBHB) candidates in \citet{Agazie+2023SMBHBs}, the nearest candidate being at $\gtrsim 95^{\circ}$ away. Future optical/infrared searches may identify additional SMBHBs in this region of the sky, and the inclusion of this MSP could therefore be valuable for targeted follow-up of new newly identified candidates as well as for improving overall PTA sensitivity to individual continuous-wave sources. 

\section{Conclusion}
\label{sec:conclusion}

We have presented the discovery and timing characterization of \source, a 3.2~ms pulsar, which was identified in a reprocessing of the GBNCC survey and independently confirmed with MeerKAT. Using observations from MeerKAT, the GBT, and Murriyang, we obtained a phase-connected timing solution spanning one year, which enabled a gamma-ray search and the subsequent discovery of gamma-ray pulsations. \source\ is in a $P_{\rm b} \approx 4.5$-day, nearly circular binary orbit with $e \approx 2 \times 10^{-5}$. Its mass function implies a minimum companion mass of 0.19\,\msun\ and a median companion mass of 0.22\,\msun\, and its location in the $P_{\rm b}-M_{\rm c}$ plane is consistent with a recycled MSP with a He-WD companion.

It has a low DM of 17.02~pc~cm$^{-3}$ and is affected by strong interstellar scintillation, causing variations in the detections of the pulsar. From the MeerKAT UHF observation we measure a scintillation bandwidth of $\Delta\nu_d=4.5(3)$~MHz, while the scintillation timescale is longer than the 10-min observation and can only be constrained to $\Delta t_d\gtrsim10$~min. The Parkes observations independently show strong epoch-to-epoch flux-density variations, with a modulation index of $m\sim1$. These variations, the low mean flux density, and the offset from the HTRU-S HighLat beam center, provide a natural explanation for its absence from previous surveys. The detection in the original GBNCC observation demonstrates the value of reprocessing archival survey data with improved search and candidate-selection techniques.

\source\ has a narrow pulse profile, with a duty cycle of only $\sim 6.5\%$ at $\sim 940$~MHz, making it potentially useful for high-precision timing. The present data yield a weighted timing residual rms of $10.2~\mu$s, while the MeerKAT observation indicates that smaller ToA uncertainties can be achieved during favorable scintillation states. However, its strong flux-density variability means that the achievable timing precision will vary considerably between epochs. Continued sensitive timing, particularly with MeerKAT, will be required to determine its long-term timing stability, measure its proper motion and hence constrain the \Pdotint, and assess its suitability for its inclusion in PTA experiments.

\section*{Acknowledgments}
L.V. and D.K. are thankful for the continuing support from the NSF Physics Frontiers Center award numbers 1430284 and 2020265. R. S. is thankful for support from the Max Planck Society. R.G. is  supported by NSF grant AST-2511757. SMR received support through the National Science Foundation Physics Frontiers Centers.
SMR is a CIFAR Fellow. 
We thank Shi Dai for allocating Parkes Green Time which substantially helped in timing the pulsar presented in this work. The CSIRO's Parkes/Murriyang radio telescope is part of the Australia Telescope National Facility (ATNF) which is funded by the Australian Government for operation as a National Facility managed by CSIRO (the Commonwealth Scientific and Industrial Research Organisation). We acknowledge the Wiradjuri people as the Traditional Owners of the Observatory site.
The MeerKAT telescope is operated by the South African Radio Astronomy Observatory, which is a facility of the National Research Foundation, an agency of the Department of Science and Innovation. SARAO acknowledges the ongoing advice and calibration of GPS systems by the National Metrology Institute of South Africa (NMISA) and the time space reference systems department of the Paris Observatory. 
TRAPUM observations used the FBFUSE and APSUSE computing clusters for data acquisition, storage and analysis. These clusters were funded and installed by the Max-Planck-Institut für Radioastronomie (MPIfR) and the Max-
Planck-Gesellschaft. FBFUSE performs beamforming operations in real-time using the \texttt{mosaic}\footnote{\url{https://github.com/wchenastro/Mosaic}} software stack \citep{Chen+2021}.
The Green Bank Observatory is a facility of the National Science Foundation operated under cooperative agreement by Associated Universities, Inc.
The National Radio Astronomy Observatory (NRAO) is a facility of the National Science Foundation operated under cooperative agreement by Associated Universities, Inc. 
The \textit{Fermi} LAT Collaboration acknowledges generous ongoing support from a number of agencies and institutes that have supported both the development and the operation of the  LAT  as  well  as  scientific  data  analysis.  These  include  the National  Aeronautics  and Space   Administration   and   the   Department   of   Energy   in the   United   States,   the Commissariat \`{a} l'Energie Atomique and the Centre National de la Recherche Scientifique /  Institut  National  de  Physique  Nucl\'{e}aire et  de  Physique  des  Particules  in  France,  the Agenzia  Spaziale  Italiana and the Istituto  Nazionale  di  Fisica  Nucleare  in  Italy,  the Ministry  of  Education,  Culture, Sports,  Science  and  Technology  (MEXT),  High  Energy Accelerator  Research Organization  (KEK)  and  Japan  Aerospace  Exploration  Agency (JAXA)  in  Japan, and  the  K.  A.  Wallenberg  Foundation,  the  Swedish  Research  Council and the Swedish National Space Board in Sweden.

The Parkes data products can be accessed through the Parkes Pulsar Data Archive (PPDA: https://data.csiro.au/domain/atnf) under project codes PX137 and P1402.

\vspace{5mm}
\facilities{Green Bank Telescope, MeerKAT telescope, Parkes/Murriyang radio telescope}

\software{astropy \citep{Astropy2013,Astropy2018}, Fermitools, PINT \citep{pint_21}, PRESTO \citep{Ransom2001}, PSRCHIVE \citep{Hotan+2004, psrchive_12}, Pulsar Survey Scraper \citep{Kaplan2022}, TEMPO2 \citep{tempo2_06}, PyGEDM \citep{Price+2021}, SeeKAT \citep{SeeKAT}}

\bibliography{gbncc_msp}{}

@ARTICLE{Price+2021,
       author = {{Price}, D.~C. and {Flynn}, C. and {Deller}, A.},
        title = "{A comparison of Galactic electron density models using PyGEDM}",
      journal = {\pasa},
         year = 2021,
        month = aug,
       volume = {38},
          eid = {e038},
        pages = {e038},
          doi = {10.1017/pasa.2021.33},
archivePrefix = {arXiv},
       eprint = {2106.15816},
 primaryClass = {astro-ph.GA},
       adsurl = {https://ui.adsabs.harvard.edu/abs/2021PASA...38...38P}
}

@ARTICLE{tempo2_06,
       author = {{Edwards}, R.~T. and {Hobbs}, G.~B. and {Manchester}, R.~N.},
        title = "{TEMPO2, a new pulsar timing package - II. The timing model and precision estimates}",
      journal = {\mnras},
         year = 2006,
        month = nov,
       volume = {372},
       number = {4},
        pages = {1549-1574},
          doi = {10.1111/j.1365-2966.2006.10870.x},
archivePrefix = {arXiv},
       eprint = {astro-ph/0607664},
 primaryClass = {astro-ph},
       adsurl = {https://ui.adsabs.harvard.edu/abs/2006MNRAS.372.1549E}
}

@ARTICLE{pint_21,
       author = {{Luo}, Jing and {Ransom}, Scott and {Demorest}, Paul and {Ray}, Paul S. and {Archibald}, Anne and {Kerr}, Matthew and {Jennings}, Ross J. and {Bachetti}, Matteo and {van Haasteren}, Rutger and {Champagne}, Chloe A. and {Colen}, Jonathan and {Phillips}, Camryn and {Zimmerman}, Josef and {Stovall}, Kevin and {Lam}, Michael T. and {Jenet}, Fredrick A.},
        title = "{PINT: A Modern Software Package for Pulsar Timing}",
      journal = {\apj},
         year = 2021,
        month = apr,
       volume = {911},
       number = {1},
          eid = {45},
        pages = {45},
          doi = {10.3847/1538-4357/abe62f},
archivePrefix = {arXiv},
       eprint = {2012.00074},
 primaryClass = {astro-ph.IM},
       adsurl = {https://ui.adsabs.harvard.edu/abs/2021ApJ...911...45L}
}

@ARTICLE{psrchive_12,
       author = {{van Straten}, Willem and {Demorest}, Paul and {Oslowski}, Stefan},
        title = "{Pulsar Data Analysis with PSRCHIVE}",
      journal = {Astronomical Research and Technology},
         year = 2012,
        month = jul,
       volume = {9},
       number = {3},
        pages = {237-256},
          doi = {10.48550/arXiv.1205.6276},
archivePrefix = {arXiv},
       eprint = {1205.6276},
 primaryClass = {astro-ph.IM},
       adsurl = {https://ui.adsabs.harvard.edu/abs/2012AR&T....9..237V}
}

@ARTICLE{Hobbs+2020,
       author = {{Hobbs}, George and {Manchester}, Richard N. and {Dunning}, Alex and {Jameson}, Andrew and {Roberts}, Paul and {George}, Daniel and {Green}, J.~A. and {Tuthill}, John and {Toomey}, Lawrence and {Kaczmarek}, Jane F. and {Mader}, Stacy and {Marquarding}, Malte and {Ahmed}, Azeem and {Amy}, Shaun W. and {Bailes}, Matthew and {Beresford}, Ron and {Bhat}, N.~D.~R. and {Bock}, Douglas C. -J. and {Bourne}, Michael and {Bowen}, Mark and {Brothers}, Michael and {Cameron}, Andrew D. and {Carretti}, Ettore and {Carter}, Nick and {Castillo}, Santy and {Chekkala}, Raji and {Cheng}, Wan and {Chung}, Yoon and {Craig}, Daniel A. and {Dai}, Shi and {Dawson}, Joanne and {Dempsey}, James and {Doherty}, Paul and {Dong}, Bin and {Edwards}, Philip and {Ergesh}, Tuohutinuer and {Gao}, Xuyang and {Han}, JinLin and {Hayman}, Douglas and {Indermuehle}, Balthasar and {Jeganathan}, Kanapathippillai and {Johnston}, Simon and {Kanoniuk}, Henry and {Kesteven}, Michael and {Kramer}, Michael and {Leach}, Mark and {Mcintyre}, Vince and {Moss}, Vanessa and {Os{\l}owski}, Stefan and {Phillips}, Chris and {Pope}, Nathan and {Preisig}, Brett and {Price}, Daniel and {Reeves}, Ken and {Reilly}, Les and {Reynolds}, John and {Robishaw}, Tim and {Roush}, Peter and {Ruckley}, Tim and {Sadler}, Elaine and {Sarkissian}, John and {Severs}, Sean and {Shannon}, Ryan and {Smart}, Ken and {Smith}, Malcolm and {Smith}, Stephanie and {Sobey}, Charlotte and {Staveley-Smith}, Lister and {Tzioumis}, Anastasios and {van Straten}, Willem and {Wang}, Nina and {Wen}, Linqing and {Whiting}, Matthew},
        title = "{An ultra-wide bandwidth (704 to 4 032 MHz) receiver for the Parkes radio telescope}",
      journal = {\pasa},
         year = 2020,
        month = apr,
       volume = {37},
          eid = {e012},
        pages = {e012},
          doi = {10.1017/pasa.2020.2},
archivePrefix = {arXiv},
       eprint = {1911.00656},
 primaryClass = {astro-ph.IM},
       adsurl = {https://ui.adsabs.harvard.edu/abs/2020PASA...37...12H}
}

@ARTICLE{Astropy2013,
       author = {{Astropy Collaboration} and {Robitaille}, Thomas P. and
         {Tollerud}, Erik J. and {Greenfield}, Perry and {Droettboom}, Michael and
         {Bray}, Erik and {Aldcroft}, Tom and {Davis}, Matt and
         {Ginsburg}, Adam and {Price-Whelan}, Adrian M. and
         {Kerzendorf}, Wolfgang E. and {Conley}, Alexander and {Crighton}, Neil and
         {Barbary}, Kyle and {Muna}, Demitri and {Ferguson}, Henry and
         {Grollier}, Fr{\'e}d{\'e}ric and {Parikh}, Madhura M. and
         {Nair}, Prasanth H. and {Unther}, Hans M. and {Deil}, Christoph and
         {Woillez}, Julien and {Conseil}, Simon and {Kramer}, Roban and
         {Turner}, James E.~H. and {Singer}, Leo and {Fox}, Ryan and
         {Weaver}, Benjamin A. and {Zabalza}, Victor and {Edwards}, Zachary I. and
         {Azalee Bostroem}, K. and {Burke}, D.~J. and {Casey}, Andrew R. and
         {Crawford}, Steven M. and {Dencheva}, Nadia and {Ely}, Justin and
         {Jenness}, Tim and {Labrie}, Kathleen and {Lim}, Pey Lian and
         {Pierfederici}, Francesco and {Pontzen}, Andrew and {Ptak}, Andy and
         {Refsdal}, Brian and {Servillat}, Mathieu and {Streicher}, Ole},
        title = "{Astropy: A community Python package for astronomy}",
      journal = {\aap},
         year = "2013",
        month = "Oct",
       volume = {558},
          eid = {A33},
        pages = {A33},
          doi = {10.1051/0004-6361/201322068},
archivePrefix = {arXiv},
       eprint = {1307.6212},
 primaryClass = {astro-ph.IM},
       adsurl = {https://ui.adsabs.harvard.edu/abs/2013A&A...558A..33A}
}

@ARTICLE{Astropy2018,
       author = {{Astropy Collaboration} and {Price-Whelan}, A.~M. and {Sip{\H{o}}cz}, B.~M. and {G{\"u}nther}, H.~M. and {Lim}, P.~L. and {Crawford}, S.~M. and {Conseil}, S. and {Shupe}, D.~L. and {Craig}, M.~W. and {Dencheva}, N. and {Ginsburg}, A. and {VanderPlas}, J.~T. and {Bradley}, L.~D. and {P{\'e}rez-Su{\'a}rez}, D. and {de Val-Borro}, M. and {Aldcroft}, T.~L. and {Cruz}, K.~L. and {Robitaille}, T.~P. and {Tollerud}, E.~J. and {Ardelean}, C. and {Babej}, T. and {Bach}, Y.~P. and {Bachetti}, M. and {Bakanov}, A.~V. and {Bamford}, S.~P. and {Barentsen}, G. and {Barmby}, P. and {Baumbach}, A. and {Berry}, K.~L. and {Biscani}, F. and {Boquien}, M. and {Bostroem}, K.~A. and {Bouma}, L.~G. and {Brammer}, G.~B. and {Bray}, E.~M. and {Breytenbach}, H. and {Buddelmeijer}, H. and {Burke}, D.~J. and {Calderone}, G. and {Cano Rodr{\'\i}guez}, J.~L. and {Cara}, M. and {Cardoso}, J.~V.~M. and {Cheedella}, S. and {Copin}, Y. and {Corrales}, L. and {Crichton}, D. and {D'Avella}, D. and {Deil}, C. and {Depagne}, {\'E}. and {Dietrich}, J.~P. and {Donath}, A. and {Droettboom}, M. and {Earl}, N. and {Erben}, T. and {Fabbro}, S. and {Ferreira}, L.~A. and {Finethy}, T. and {Fox}, R.~T. and {Garrison}, L.~H. and {Gibbons}, S.~L.~J. and {Goldstein}, D.~A. and {Gommers}, R. and {Greco}, J.~P. and {Greenfield}, P. and {Groener}, A.~M. and {Grollier}, F. and {Hagen}, A. and {Hirst}, P. and {Homeier}, D. and {Horton}, A.~J. and {Hosseinzadeh}, G. and {Hu}, L. and {Hunkeler}, J.~S. and {Ivezi{\'c}}, {\v{Z}}. and {Jain}, A. and {Jenness}, T. and {Kanarek}, G. and {Kendrew}, S. and {Kern}, N.~S. and {Kerzendorf}, W.~E. and {Khvalko}, A. and {King}, J. and {Kirkby}, D. and {Kulkarni}, A.~M. and {Kumar}, A. and {Lee}, A. and {Lenz}, D. and {Littlefair}, S.~P. and {Ma}, Z. and {Macleod}, D.~M. and {Mastropietro}, M. and {McCully}, C. and {Montagnac}, S. and {Morris}, B.~M. and {Mueller}, M. and {Mumford}, S.~J. and {Muna}, D. and {Murphy}, N.~A. and {Nelson}, S. and {Nguyen}, G.~H. and {Ninan}, J.~P. and {N{\"o}the}, M. and {Ogaz}, S. and {Oh}, S. and {Parejko}, J.~K. and {Parley}, N. and {Pascual}, S. and {Patil}, R. and {Patil}, A.~A. and {Plunkett}, A.~L. and {Prochaska}, J.~X. and {Rastogi}, T. and {Reddy Janga}, V. and {Sabater}, J. and {Sakurikar}, P. and {Seifert}, M. and {Sherbert}, L.~E. and {Sherwood-Taylor}, H. and {Shih}, A.~Y. and {Sick}, J. and {Silbiger}, M.~T. and {Singanamalla}, S. and {Singer}, L.~P. and {Sladen}, P.~H. and {Sooley}, K.~A. and {Sornarajah}, S. and {Streicher}, O. and {Teuben}, P. and {Thomas}, S.~W. and {Tremblay}, G.~R. and {Turner}, J.~E.~H. and {Terr{\'o}n}, V. and {van Kerkwijk}, M.~H. and {de la Vega}, A. and {Watkins}, L.~L. and {Weaver}, B.~A. and {Whitmore}, J.~B. and {Woillez}, J. and {Zabalza}, V. and {Astropy Contributors}},
        title = "{The Astropy Project: Building an Open-science Project and Status of the v2.0 Core Package}",
      journal = {\aj},
         year = 2018,
        month = sep,
       volume = {156},
       number = {3},
          eid = {123},
        pages = {123},
          doi = {10.3847/1538-3881/aabc4f},
archivePrefix = {arXiv},
       eprint = {1801.02634},
 primaryClass = {astro-ph.IM},
       adsurl = {https://ui.adsabs.harvard.edu/abs/2018AJ....156..123A}
}

@ARTICLE{Manchester+2005,
       author = {{Manchester}, R.~N. and {Hobbs}, G.~B. and {Teoh}, A. and {Hobbs}, M.},
        title = "{The Australia Telescope National Facility Pulsar Catalogue}",
      journal = {\aj},
         year = 2005,
        month = apr,
       volume = {129},
       number = {4},
        pages = {1993-2006},
          doi = {10.1086/428488},
archivePrefix = {arXiv},
       eprint = {astro-ph/0412641},
 primaryClass = {astro-ph},
       adsurl = {https://ui.adsabs.harvard.edu/abs/2005AJ....129.1993M}
}

@ARTICLE{Alpar1982,
       author = {{Alpar}, M.~A. and {Cheng}, A.~F. and {Ruderman}, M.~A. and {Shaham}, J.},
        title = "{A new class of radio pulsars}",
      journal = {\nat},
         year = 1982,
        month = dec,
       volume = {300},
       number = {5894},
        pages = {728-730},
          doi = {10.1038/300728a0},
       adsurl = {https://ui.adsabs.harvard.edu/abs/1982Natur.300..728A}
}

@ARTICLE{Radhakrishnan+Srinivasan1982,
       author = {{Radhakrishnan}, V. and {Srinivasan}, G.},
        title = "{On the origin of the recently discovered ultra-rapid pulsar}",
      journal = {Current Science},
         year = 1982,
        month = dec,
       volume = {51},
        pages = {1096-1099},
       adsurl = {https://ui.adsabs.harvard.edu/abs/1982CSci...51.1096R}
}

@ARTICLE{Hellings+Downs1983,
       author = {{Hellings}, R.~W. and {Downs}, G.~S.},
        title = "{Upper limits on the isotropic gravitational radiation background from pulsar timing analysis.}",
      journal = {\apjl},
         year = 1983,
        month = feb,
       volume = {265},
        pages = {L39-L42},
          doi = {10.1086/183954},
       adsurl = {https://ui.adsabs.harvard.edu/abs/1983ApJ...265L..39H}
}

@ARTICLE{Foster+Backer1990,
       author = {{Foster}, R.~S. and {Backer}, D.~C.},
        title = "{Constructing a Pulsar Timing Array}",
      journal = {\apj},
         year = 1990,
        month = sep,
       volume = {361},
        pages = {300},
          doi = {10.1086/169195},
       adsurl = {https://ui.adsabs.harvard.edu/abs/1990ApJ...361..300F}
}

@ARTICLE{Nanograv+23,
       author = {{Agazie}, Gabriella and {Anumarlapudi}, Akash and {Archibald}, Anne M. and {Arzoumanian}, Zaven and {Baker}, Paul T. and {B{\'e}csy}, Bence and {Blecha}, Laura and {Brazier}, Adam and {Brook}, Paul R. and {Burke-Spolaor}, Sarah and {Burnette}, Rand and {Case}, Robin and {Charisi}, Maria and {Chatterjee}, Shami and {Chatziioannou}, Katerina and {Cheeseboro}, Belinda D. and {Chen}, Siyuan and {Cohen}, Tyler and {Cordes}, James M. and {Cornish}, Neil J. and {Crawford}, Fronefield and {Cromartie}, H. Thankful and {Crowter}, Kathryn and {Cutler}, Curt J. and {Decesar}, Megan E. and {Degan}, Dallas and {Demorest}, Paul B. and {Deng}, Heling and {Dolch}, Timothy and {Drachler}, Brendan and {Ellis}, Justin A. and {Ferrara}, Elizabeth C. and {Fiore}, William and {Fonseca}, Emmanuel and {Freedman}, Gabriel E. and {Garver-Daniels}, Nate and {Gentile}, Peter A. and {Gersbach}, Kyle A. and {Glaser}, Joseph and {Good}, Deborah C. and {G{\"u}ltekin}, Kayhan and {Hazboun}, Jeffrey S. and {Hourihane}, Sophie and {Islo}, Kristina and {Jennings}, Ross J. and {Johnson}, Aaron D. and {Jones}, Megan L. and {Kaiser}, Andrew R. and {Kaplan}, David L. and {Kelley}, Luke Zoltan and {Kerr}, Matthew and {Key}, Joey S. and {Klein}, Tonia C. and {Laal}, Nima and {Lam}, Michael T. and {Lamb}, William G. and {Lazio}, T. Joseph W. and {Lewandowska}, Natalia and {Littenberg}, Tyson B. and {Liu}, Tingting and {Lommen}, Andrea and {Lorimer}, Duncan R. and {Luo}, Jing and {Lynch}, Ryan S. and {Ma}, Chung-Pei and {Madison}, Dustin R. and {Mattson}, Margaret A. and {McEwen}, Alexander and {McKee}, James W. and {McLaughlin}, Maura A. and {McMann}, Natasha and {Meyers}, Bradley W. and {Meyers}, Patrick M. and {Mingarelli}, Chiara M.~F. and {Mitridate}, Andrea and {Natarajan}, Priyamvada and {Ng}, Cherry and {Nice}, David J. and {Ocker}, Stella Koch and {Olum}, Ken D. and {Pennucci}, Timothy T. and {Perera}, Benetge B.~P. and {Petrov}, Polina and {Pol}, Nihan S. and {Radovan}, Henri A. and {Ransom}, Scott M. and {Ray}, Paul S. and {Romano}, Joseph D. and {Sardesai}, Shashwat C. and {Schmiedekamp}, Ann and {Schmiedekamp}, Carl and {Schmitz}, Kai and {Schult}, Levi and {Shapiro-Albert}, Brent J. and {Siemens}, Xavier and {Simon}, Joseph and {Siwek}, Magdalena S. and {Stairs}, Ingrid H. and {Stinebring}, Daniel R. and {Stovall}, Kevin and {Sun}, Jerry P. and {Susobhanan}, Abhimanyu and {Swiggum}, Joseph K. and {Taylor}, Jacob and {Taylor}, Stephen R. and {Turner}, Jacob E. and {Unal}, Caner and {Vallisneri}, Michele and {van Haasteren}, Rutger and {Vigeland}, Sarah J. and {Wahl}, Haley M. and {Wang}, Qiaohong and {Witt}, Caitlin A. and {Young}, Olivia and {Nanograv Collaboration}},
        title = "{The NANOGrav 15 yr Data Set: Evidence for a Gravitational-wave Background}",
      journal = {\apjl},
         year = 2023,
        month = jul,
       volume = {951},
       number = {1},
          eid = {L8},
        pages = {L8},
          doi = {10.3847/2041-8213/acdac6},
archivePrefix = {arXiv},
       eprint = {2306.16213},
 primaryClass = {astro-ph.HE},
       adsurl = {https://ui.adsabs.harvard.edu/abs/2023ApJ...951L...8A}
}

@ARTICLE{Lam2018,
       author = {{Lam}, M.~T.},
        title = "{Optimizing Pulsar Timing Array Observational Cadences for Sensitivity to Low-frequency Gravitational-wave Sources}",
      journal = {\apj},
         year = 2018,
        month = nov,
       volume = {868},
       number = {1},
          eid = {33},
        pages = {33},
          doi = {10.3847/1538-4357/aae533},
archivePrefix = {arXiv},
       eprint = {1808.10071},
 primaryClass = {astro-ph.HE},
       adsurl = {https://ui.adsabs.harvard.edu/abs/2018ApJ...868...33L}
}

@ARTICLE{Miles+2025,
       author = {{Miles}, Matthew T. and {Shannon}, Ryan M. and {Reardon}, Daniel J. and {Bailes}, Matthew and {Champion}, David J. and {Geyer}, Marisa and {Gitika}, Pratyasha and {Grunthal}, Kathrin and {Keith}, Michael J. and {Kramer}, Michael and {Kulkarni}, Atharva D. and {Nathan}, Rowina S. and {Parthasarathy}, Aditya and {Singha}, Jaikhomba and {Theureau}, Gilles and {Thrane}, Eric and {Abbate}, Federico and {Buchner}, Sarah and {Cameron}, Andrew D. and {Camilo}, Fernando and {Moreschi}, Beatrice E. and {Shaifullah}, Golam and {Shamohammadi}, Mohsen and {Possenti}, Andrea and {Krishnan}, Vivek Venkatraman},
        title = "{The MeerKAT Pulsar Timing Array: the first search for gravitational waves with the MeerKAT radio telescope}",
      journal = {\mnras},
         year = 2025,
        month = jan,
       volume = {536},
       number = {2},
        pages = {1489-1500},
          doi = {10.1093/mnras/stae2571},
archivePrefix = {arXiv},
       eprint = {2412.01153},
 primaryClass = {astro-ph.HE},
       adsurl = {https://ui.adsabs.harvard.edu/abs/2025MNRAS.536.1489M}
}

@ARTICLE{Siemens+2013,
       author = {{Siemens}, Xavier and {Ellis}, Justin and {Jenet}, Fredrick and {Romano}, Joseph D.},
        title = "{The stochastic background: scaling laws and time to detection for pulsar timing arrays}",
      journal = {Classical and Quantum Gravity},
         year = 2013,
        month = nov,
       volume = {30},
       number = {22},
          eid = {224015},
        pages = {224015},
          doi = {10.1088/0264-9381/30/22/224015},
archivePrefix = {arXiv},
       eprint = {1305.3196},
 primaryClass = {astro-ph.IM},
       adsurl = {https://ui.adsabs.harvard.edu/abs/2013CQGra..30v4015S}
}

@ARTICLE{Cordes+Shannon2010,
       author = {{Cordes}, J.~M. and {Shannon}, R.~M.},
        title = "{A Measurement Model for Precision Pulsar Timing}",
      journal = {arXiv e-prints},
         year = 2010,
        month = oct,
          eid = {arXiv:1010.3785},
        pages = {arXiv:1010.3785},
          doi = {10.48550/arXiv.1010.3785},
archivePrefix = {arXiv},
       eprint = {1010.3785},
 primaryClass = {astro-ph.IM},
       adsurl = {https://ui.adsabs.harvard.edu/abs/2010arXiv1010.3785C}
}

@ARTICLE{Keith+2013,
       author = {{Keith}, M.~J. and {Coles}, W. and {Shannon}, R.~M. and {Hobbs}, G.~B. and {Manchester}, R.~N. and {Bailes}, M. and {Bhat}, N.~D.~R. and {Burke-Spolaor}, S. and {Champion}, D.~J. and {Chaudhary}, A. and {Hotan}, A.~W. and {Khoo}, J. and {Kocz}, J. and {Os{\l}owski}, S. and {Ravi}, V. and {Reynolds}, J.~E. and {Sarkissian}, J. and {van Straten}, W. and {Yardley}, D.~R.~B.},
        title = "{Measurement and correction of variations in interstellar dispersion in high-precision pulsar timing}",
      journal = {\mnras},
         year = 2013,
        month = mar,
       volume = {429},
       number = {3},
        pages = {2161-2174},
          doi = {10.1093/mnras/sts486},
archivePrefix = {arXiv},
       eprint = {1211.5887},
 primaryClass = {astro-ph.GA},
       adsurl = {https://ui.adsabs.harvard.edu/abs/2013MNRAS.429.2161K}
}

@ARTICLE{Reardon+2020,
       author = {{Reardon}, Daniel J. and {Coles}, William A. and {Bailes}, Matthew and {Bhat}, N.~D. Ramesh and {Dai}, Shi and {Hobbs}, George B. and {Kerr}, Matthew and {Manchester}, Richard N. and {Os{\l}owski}, Stefan and {Parthasarathy}, Aditya and {Russell}, Christopher J. and {Shannon}, Ryan M. and {Spiewak}, Ren{\'e}e and {Toomey}, Lawrence and {Tuntsov}, Artem V. and {van Straten}, Willem and {Walker}, Mark A. and {Wang}, Jingbo and {Zhang}, Lei and {Zhu}, Xing-Jiang},
        title = "{Precision Orbital Dynamics from Interstellar Scintillation Arcs for PSR J0437-4715}",
      journal = {\apj},
         year = 2020,
        month = dec,
       volume = {904},
       number = {2},
          eid = {104},
        pages = {104},
          doi = {10.3847/1538-4357/abbd40},
archivePrefix = {arXiv},
       eprint = {2009.12757},
 primaryClass = {astro-ph.HE},
       adsurl = {https://ui.adsabs.harvard.edu/abs/2020ApJ...904..104R}
}

@inproceedings{Stappers+Kramer2018,
  author = "Stappers, Ben  and  Kramer, Michael",
  title = "{An Update on TRAPUM}",
  doi = "10.22323/1.277.0009",
  booktitle = "Proceedings of MeerKAT Science: On the Pathway to the SKA {\textemdash} PoS(MeerKAT2016)",
  year = 2018,
  volume = "277",
  pages = "009",
  adsurl = {https://ui.adsabs.harvard.edu/abs/2016mks..confE...9S}
}

@ARTICLE{Tauris+2017,
       author = {{Tauris}, T.~M. and {Kramer}, M. and {Freire}, P.~C.~C. and {Wex}, N. and {Janka}, H. -T. and {Langer}, N. and {Podsiadlowski}, Ph. and {Bozzo}, E. and {Chaty}, S. and {Kruckow}, M.~U. and {van den Heuvel}, E.~P.~J. and {Antoniadis}, J. and {Breton}, R.~P. and {Champion}, D.~J.},
        title = "{Formation of Double Neutron Star Systems}",
      journal = {\apj},
         year = 2017,
        month = sep,
       volume = {846},
       number = {2},
          eid = {170},
        pages = {170},
          doi = {10.3847/1538-4357/aa7e89},
archivePrefix = {arXiv},
       eprint = {1706.09438},
 primaryClass = {astro-ph.HE},
       adsurl = {https://ui.adsabs.harvard.edu/abs/2017ApJ...846..170T}
}

@ARTICLE{Voisin+2020,
       author = {{Voisin}, G. and {Cognard}, I. and {Freire}, P.~C.~C. and {Wex}, N. and {Guillemot}, L. and {Desvignes}, G. and {Kramer}, M. and {Theureau}, G.},
        title = "{An improved test of the strong equivalence principle with the pulsar in a triple star system}",
      journal = {\aap},
         year = 2020,
        month = jun,
       volume = {638},
          eid = {A24},
        pages = {A24},
          doi = {10.1051/0004-6361/202038104},
archivePrefix = {arXiv},
       eprint = {2005.01388},
 primaryClass = {gr-qc},
       adsurl = {https://ui.adsabs.harvard.edu/abs/2020A&A...638A..24V}
}

@ARTICLE{Kramer+2021,
       author = {{Kramer}, M. and {Stairs}, I.~H. and {Manchester}, R.~N. and {Wex}, N. and {Deller}, A.~T. and {Coles}, W.~A. and {Ali}, M. and {Burgay}, M. and {Camilo}, F. and {Cognard}, I. and {Damour}, T. and {Desvignes}, G. and {Ferdman}, R.~D. and {Freire}, P.~C.~C. and {Grondin}, S. and {Guillemot}, L. and {Hobbs}, G.~B. and {Janssen}, G. and {Karuppusamy}, R. and {Lorimer}, D.~R. and {Lyne}, A.~G. and {McKee}, J.~W. and {McLaughlin}, M. and {M{\"u}nch}, L.~E. and {Perera}, B.~B.~P. and {Pol}, N. and {Possenti}, A. and {Sarkissian}, J. and {Stappers}, B.~W. and {Theureau}, G.},
        title = "{Strong-Field Gravity Tests with the Double Pulsar}",
      journal = {Physical Review X},
         year = 2021,
        month = oct,
       volume = {11},
       number = {4},
          eid = {041050},
        pages = {041050},
          doi = {10.1103/PhysRevX.11.041050},
archivePrefix = {arXiv},
       eprint = {2112.06795},
 primaryClass = {astro-ph.HE},
       adsurl = {https://ui.adsabs.harvard.edu/abs/2021PhRvX..11d1050K}
}

@ARTICLE{Reardon+2023,
       author = {{Reardon}, Daniel J. and {Zic}, Andrew and {Shannon}, Ryan M. and {Hobbs}, George B. and {Bailes}, Matthew and {Di Marco}, Valentina and {Kapur}, Agastya and {Rogers}, Axl F. and {Thrane}, Eric and {Askew}, Jacob and {Bhat}, N.~D. Ramesh and {Cameron}, Andrew and {Cury{\l}o}, Ma{\l}gorzata and {Coles}, William A. and {Dai}, Shi and {Goncharov}, Boris and {Kerr}, Matthew and {Kulkarni}, Atharva and {Levin}, Yuri and {Lower}, Marcus E. and {Manchester}, Richard N. and {Mandow}, Rami and {Miles}, Matthew T. and {Nathan}, Rowina S. and {Os{\l}owski}, Stefan and {Russell}, Christopher J. and {Spiewak}, Ren{\'e}e and {Zhang}, Songbo and {Zhu}, Xing-Jiang},
        title = "{Search for an Isotropic Gravitational-wave Background with the Parkes Pulsar Timing Array}",
      journal = {\apjl},
         year = 2023,
        month = jul,
       volume = {951},
       number = {1},
          eid = {L6},
        pages = {L6},
          doi = {10.3847/2041-8213/acdd02},
archivePrefix = {arXiv},
       eprint = {2306.16215},
 primaryClass = {astro-ph.HE},
       adsurl = {https://ui.adsabs.harvard.edu/abs/2023ApJ...951L...6R}
}

@ARTICLE{Cromartie+2020,
       author = {{Cromartie}, H.~T. and {Fonseca}, E. and {Ransom}, S.~M. and {Demorest}, P.~B. and {Arzoumanian}, Z. and {Blumer}, H. and {Brook}, P.~R. and {DeCesar}, M.~E. and {Dolch}, T. and {Ellis}, J.~A. and {Ferdman}, R.~D. and {Ferrara}, E.~C. and {Garver-Daniels}, N. and {Gentile}, P.~A. and {Jones}, M.~L. and {Lam}, M.~T. and {Lorimer}, D.~R. and {Lynch}, R.~S. and {McLaughlin}, M.~A. and {Ng}, C. and {Nice}, D.~J. and {Pennucci}, T.~T. and {Spiewak}, R. and {Stairs}, I.~H. and {Stovall}, K. and {Swiggum}, J.~K. and {Zhu}, W.~W.},
        title = "{Relativistic Shapiro delay measurements of an extremely massive millisecond pulsar}",
      journal = {Nature Astronomy},
         year = 2020,
        month = jan,
       volume = {4},
        pages = {72-76},
          doi = {10.1038/s41550-019-0880-2},
archivePrefix = {arXiv},
       eprint = {1904.06759},
 primaryClass = {astro-ph.HE},
       adsurl = {https://ui.adsabs.harvard.edu/abs/2020NatAs...4...72C}
}

@ARTICLE{Sengar+2023,
       author = {{Sengar}, R. and {Bailes}, M. and {Balakrishnan}, V. and {Bernadich}, M.~C. i. and {Burgay}, M. and {Barr}, E.~D. and {Flynn}, C.~M.~L. and {Stevenson}, R. Shannon S. and {Wongphechauxsorn}, J.},
        title = "{Discovery of 37 new pulsars through GPU-accelerated reprocessing of archival data of the Parkes multibeam pulsar survey}",
      journal = {\mnras},
         year = 2023,
        month = jun,
       volume = {522},
       number = {1},
        pages = {1071-1090},
          doi = {10.1093/mnras/stad508},
archivePrefix = {arXiv},
       eprint = {2302.00255},
 primaryClass = {astro-ph.HE},
       adsurl = {https://ui.adsabs.harvard.edu/abs/2023MNRAS.522.1071S}
}

@ARTICLE{Sengar+2025,
       author = {{Sengar}, R. and {Bailes}, M. and {Balakrishnan}, V. and {Barr}, E.~D. and {Bhat}, N.~D.~R. and {Burgay}, M. and {Bernadich}, M.~C.~I. and {Cameron}, A.~D. and {Champion}, D.~J. and {Chen}, W. and {Flynn}, C.~M.~L. and {Jameson}, A. and {Johnston}, S. and {Keith}, M.~J. and {Kramer}, M. and {Morello}, V. and {Ng}, C. and {Possenti}, A. and {Stevenson}, S. and {Shannon}, R.~M. and {van Straten}, W. and {Wongphechauxsorn}, J.},
        title = "{The High Time Resolution Universe Pulsar Survey - XIX. A coherent GPU-accelerated reprocessing and the discovery of 71 pulsars in the Southern Galactic plane}",
      journal = {\mnras},
         year = 2025,
        month = feb,
       volume = {536},
       number = {4},
        pages = {3159-3176},
          doi = {10.1093/mnras/stae2716},
archivePrefix = {arXiv},
       eprint = {2412.07104},
 primaryClass = {astro-ph.HE},
       adsurl = {https://ui.adsabs.harvard.edu/abs/2025MNRAS.536.3159S}
}

@ARTICLE{Stovall+2014,
       author = {{Stovall}, K. and {Lynch}, R.~S. and {Ransom}, S.~M. and {Archibald}, A.~M. and {Banaszak}, S. and {Biwer}, C.~M. and {Boyles}, J. and {Dartez}, L.~P. and {Day}, D. and {Ford}, A.~J. and {Flanigan}, J. and {Garcia}, A. and {Hessels}, J.~W.~T. and {Hinojosa}, J. and {Jenet}, F.~A. and {Kaplan}, D.~L. and {Karako-Argaman}, C. and {Kaspi}, V.~M. and {Kondratiev}, V.~I. and {Leake}, S. and {Lorimer}, D.~R. and {Lunsford}, G. and {Martinez}, J.~G. and {Mata}, A. and {McLaughlin}, M.~A. and {Roberts}, M.~S.~E. and {Rohr}, M.~D. and {Siemens}, X. and {Stairs}, I.~H. and {van Leeuwen}, J. and {Walker}, A.~N. and {Wells}, B.~L.},
        title = "{The Green Bank Northern Celestial Cap Pulsar Survey. I. Survey Description, Data Analysis, and Initial Results}",
      journal = {\apj},
         year = 2014,
        month = aug,
       volume = {791},
       number = {1},
          eid = {67},
        pages = {67},
          doi = {10.1088/0004-637X/791/1/67},
archivePrefix = {arXiv},
       eprint = {1406.5214},
 primaryClass = {astro-ph.HE},
       adsurl = {https://ui.adsabs.harvard.edu/abs/2014ApJ...791...67S}
}

@ARTICLE{Gitika+2023,
       author = {{Gitika}, P. and {Bailes}, M. and {Shannon}, R.~M. and {Reardon}, D.~J. and {Cameron}, A.~D. and {Shamohammadi}, M. and {Miles}, M.~T. and {Flynn}, C.~M.~L. and {Corongiu}, A. and {Kramer}, M.},
        title = "{Flux density monitoring of 89 millisecond pulsars with MeerKAT}",
      journal = {\mnras},
         year = 2023,
        month = dec,
       volume = {526},
       number = {3},
        pages = {3370-3385},
          doi = {10.1093/mnras/stad2841},
archivePrefix = {arXiv},
       eprint = {2309.07564},
 primaryClass = {astro-ph.HE},
       adsurl = {https://ui.adsabs.harvard.edu/abs/2023MNRAS.526.3370G}
}

@ARTICLE{Backer+1982,
       author = {{Backer}, D.~C. and {Kulkarni}, S.~R. and {Heiles}, C. and {Davis}, M.~M. and {Goss}, W.~M.},
        title = "{A millisecond pulsar}",
      journal = {\nat},
         year = 1982,
        month = dec,
       volume = {300},
       number = {5893},
        pages = {615-618},
          doi = {10.1038/300615a0},
       adsurl = {https://ui.adsabs.harvard.edu/abs/1982Natur.300..615B}
}

@ARTICLE{Fabian+1983,
       author = {{Fabian}, A.~C. and {Pringle}, J.~E. and {Verbunt}, F. and {Wade}, R.~A.},
        title = "{Do galactic bulge X-ray sources evolve into millisecond pulsars?}",
      journal = {\nat},
         year = 1983,
        month = jan,
       volume = {301},
       number = {5897},
        pages = {222-223},
          doi = {10.1038/301222a0},
       adsurl = {https://ui.adsabs.harvard.edu/abs/1983Natur.301..222F}
}

@ARTICLE{Thongmeearkom+2026,
       author = {{Thongmeearkom}, T. and {Clark}, C.~J. and {Breton}, R.~P. and {Burgay}, M. and {Nieder}, L. and {Dodge}, O.~G. and {McGloughlin}, B. and {Barr}, E.~D. and {Buchner}, S. and {Stappers}, B.~W. and {Berteaud}, J. and {Ferrara}, E.~C. and {Freire}, P.~C.~C. and {Levin}, L. and {Ransom}, S.~M. and {Vleeschower}, L. and {Belmonte D{\'\i}az}, S. and {Calore}, F. and {Cognard}, I. and {Dhillon}, V.~S. and {Grie{\ss}meier}, J.-M. and {Karuppusamy}, R. and {Kennedy}, M.~R. and {Kramer}, M. and {Padmanabh}, P.~V. and {Papa}, M.~A. and {Phosrisom}, A. and {Steltner}, B.},
        title = "{Pulsar discoveries from the TRAPUM UHF survey of Fermi-LAT sources}",
      journal = {\mnras},
         year = 2026,
        month = apr,
       volume = {547},
       number = {4},
          eid = {stag406},
        pages = {stag406},
          doi = {10.1093/mnras/stag406},
archivePrefix = {arXiv},
       eprint = {2602.20811},
 primaryClass = {astro-ph.HE},
       adsurl = {https://ui.adsabs.harvard.edu/abs/2026MNRAS.547ag406T}
}

@ARTICLE{Sengar+2026,
       author = {{Sengar}, R. and {Vleeschower}, L. and {Belmonte D{\'\i}az}, S. and {Calore}, F. and {Cognard}, I. and {Dhillon}, V.~S. and {Grie{\ss}meier}, J.-M. and {Karuppusamy}, R. and {Kennedy}, M.~R. and {Kramer}, M. and {Padmanabh}, P.~V. and {Papa}, M.~A. and {Phosrisom}, A. and {Steltner}, B.},
        title = "{Discovery of 51 pulsars in a reprocessing of the Green Bank North Celestial Cap Survey}",
      journal = {\apj},
         year = 2026,
        month = oct,
       volume = {547},
       number = {4},
          eid = {stag406},
        pages = {stag406},
          doi = {10.1093/mnras/stag406},
archivePrefix = {arXiv},
       eprint = {2602.20811},
 primaryClass = {astro-ph.HE},
       adsurl = {https://ui.adsabs.harvard.edu/abs/2026MNRAS.547ag406T}
}

@INPROCEEDINGS{Barr2018,
       author = {{Barr}, Ewan D.},
        title = "{An S-band Receiver and Backend System for MeerKAT}",
    booktitle = {Pulsar Astrophysics the Next Fifty Years},
         year = 2018,
       editor = {{Weltevrede}, P. and {Perera}, B.~B.~P. and {Preston}, L.~L. and {Sanidas}, S.},
       series = {IAU Symposium},
       volume = {337},
        month = aug,
        pages = {175-178},
          doi = {10.1017/S1743921317009036},
       adsurl = {https://ui.adsabs.harvard.edu/abs/2018IAUS..337..175B}
}

@ARTICLE{Chen+2021,
       author = {{Chen}, Weiwei and {Barr}, Ewan and {Karuppusamy}, Ramesh and {Kramer}, Michael and {Stappers}, Benjamin},
        title = "{Wide Field Beamformed Observation with MeerKAT}",
      journal = {Journal of Astronomical Instrumentation},
         year = 2021,
        month = jan,
       volume = {10},
       number = {3},
          eid = {2150013-178},
        pages = {2150013-178},
          doi = {10.1142/S2251171721500136},
archivePrefix = {arXiv},
       eprint = {2110.01667},
 primaryClass = {astro-ph.IM},
       adsurl = {https://ui.adsabs.harvard.edu/abs/2021JAI....1050013C}
}

@ARTICLE{Clark+2023,
       author = {{Clark}, C.~J. and {Breton}, R.~P. and {Barr}, E.~D. and {Burgay}, M. and {Thongmeearkom}, T. and {Nieder}, L. and {Buchner}, S. and {Stappers}, B. and {Kramer}, M. and {Becker}, W. and {Mayer}, M. and {Phosrisom}, A. and {Ashok}, A. and {Bezuidenhout}, M.~C. and {Calore}, F. and {Cognard}, I. and {Freire}, P.~C.~C. and {Geyer}, M. and {Grie{\ss}meier}, J.-M. and {Karuppusamy}, R. and {Levin}, L. and {Padmanabh}, P.~V. and {Possenti}, A. and {Ransom}, S. and {Serylak}, M. and {Venkatraman Krishnan}, V. and {Vleeschower}, L. and {Behrend}, J. and {Champion}, D.~J. and {Chen}, W. and {Horn}, D. and {Keane}, E.~F. and {K{\"u}nkel}, L. and {Men}, Y. and {Ridolfi}, A. and {Dhillon}, V.~S. and {Marsh}, T.~R. and {Papa}, M.~A.},
        title = "{The TRAPUM L-band survey for pulsars in Fermi-LAT gamma-ray sources}",
      journal = {\mnras},
         year = 2023,
        month = mar,
       volume = {519},
       number = {4},
        pages = {5590-5606},
          doi = {10.1093/mnras/stac3742},
archivePrefix = {arXiv},
       eprint = {2212.08528},
 primaryClass = {astro-ph.HE},
       adsurl = {https://ui.adsabs.harvard.edu/abs/2023MNRAS.519.5590C}
}

@ARTICLE{vanStraten+2011,
       author = {{van Straten}, W. and {Bailes}, M.},
        title = "{DSPSR: Digital Signal Processing Software for Pulsar Astronomy}",
      journal = {\pasa},
         year = 2011,
        month = jan,
       volume = {28},
       number = {1},
        pages = {1-14},
          doi = {10.1071/AS10021},
archivePrefix = {arXiv},
       eprint = {1008.3973},
 primaryClass = {astro-ph.IM},
       adsurl = {https://ui.adsabs.harvard.edu/abs/2011PASA...28....1V}
}

@ARTICLE{Morello+2019,
       author = {{Morello}, V. and {Barr}, E.~D. and {Cooper}, S. and {Bailes}, M. and {Bates}, S. and {Bhat}, N.~D.~R. and {Burgay}, M. and {Burke-Spolaor}, S. and {Cameron}, A.~D. and {Champion}, D.~J. and {Eatough}, R.~P. and {Flynn}, C.~M.~L. and {Jameson}, A. and {Johnston}, S. and {Keith}, M.~J. and {Keane}, E.~F. and {Kramer}, M. and {Levin}, L. and {Ng}, C. and {Petroff}, E. and {Possenti}, A. and {Stappers}, B.~W. and {van Straten}, W. and {Tiburzi}, C.},
        title = "{The High Time Resolution Universe survey - XIV. Discovery of 23 pulsars through GPU-accelerated reprocessing}",
      journal = {\mnras},
         year = 2019,
        month = mar,
       volume = {483},
       number = {3},
        pages = {3673-3685},
          doi = {10.1093/mnras/sty3328},
archivePrefix = {arXiv},
       eprint = {1811.04929},
 primaryClass = {astro-ph.IM},
       adsurl = {https://ui.adsabs.harvard.edu/abs/2019MNRAS.483.3673M}
}

@ARTICLE{Hotan+2004,
       author = {{Hotan}, A.~W. and {van Straten}, W. and {Manchester}, R.~N.},
        title = "{PSRCHIVE and PSRFITS: An Open Approach to Radio Pulsar Data Storage and Analysis}",
      journal = {\pasa},
         year = 2004,
        month = jan,
       volume = {21},
       number = {3},
        pages = {302-309},
          doi = {10.1071/AS04022},
archivePrefix = {arXiv},
       eprint = {astro-ph/0404549},
 primaryClass = {astro-ph},
       adsurl = {https://ui.adsabs.harvard.edu/abs/2004PASA...21..302H}
}

@software{Kaplan2022,
       author = {{Kaplan}, David L.},
        title = "{PSS: Pulsar Survey Scraper}",
 howpublished = {Astrophysics Source Code Library, record ascl:2210.001},
         year = 2022,
        month = oct,
          eid = {ascl:2210.001},
archivePrefix = {ascl},
       eprint = {2210.001},
       adsurl = {https://ui.adsabs.harvard.edu/abs/2022ascl.soft10001K}
}

@ARTICLE{Freire+2001,
       author = {{Freire}, P.~C. and {Kramer}, M. and {Lyne}, A.~G.},
        title = "{Determination of the orbital parameters of binary pulsars}",
      journal = {\mnras},
         year = 2001,
        month = apr,
       volume = {322},
       number = {4},
        pages = {885-890},
          doi = {10.1046/j.1365-8711.2001.04200.x},
archivePrefix = {arXiv},
       eprint = {astro-ph/0010463},
 primaryClass = {astro-ph},
       adsurl = {https://ui.adsabs.harvard.edu/abs/2001MNRAS.322..885F}
}

@ARTICLE{Folkner+2009,
       author = {{Folkner}, W.~M. and {Williams}, J.~G. and {Boggs}, D.~H.},
        title = "{The Planetary and Lunar Ephemeris DE 421}",
      journal = {Interplanetary Network Progress Report},
         year = 2009,
        month = aug,
       volume = {42-178},
        pages = {1-34},
       adsurl = {https://ui.adsabs.harvard.edu/abs/2009IPNPR.178C...1F}
}

@ARTICLE{Hobbs+2006,
       author = {{Hobbs}, G. and {Edwards}, R. and {Manchester}, R.},
        title = "{TEMPO2: a New Pulsar Timing Package}",
      journal = {Chinese Journal of Astronomy and Astrophysics Supplement},
         year = 2006,
        month = dec,
       volume = {6},
       number = {S2},
        pages = {189-192},
       adsurl = {https://ui.adsabs.harvard.edu/abs/2006ChJAS...6b.189H}
}

@ARTICLE{Edwards+2006,
       author = {{Edwards}, R.~T. and {Hobbs}, G.~B. and {Manchester}, R.~N.},
        title = "{TEMPO2, a new pulsar timing package - II. The timing model and precision estimates}",
      journal = {\mnras},
         year = 2006,
        month = nov,
       volume = {372},
       number = {4},
        pages = {1549-1574},
          doi = {10.1111/j.1365-2966.2006.10870.x},
archivePrefix = {arXiv},
       eprint = {astro-ph/0607664},
 primaryClass = {astro-ph},
       adsurl = {https://ui.adsabs.harvard.edu/abs/2006MNRAS.372.1549E}
}

@article{Lange+2001,
    author = {Lange, Ch. and Camilo, F. and Wex, N. and Kramer, M. and Backer, D.C. and Lyne, A.G. and Doroshenko, O.},
    title = {Precision timing measurements of PSR J1012+5307},
    journal = {Monthly Notices of the Royal Astronomical Society},
    volume = {326},
    number = {1},
    pages = {274-282},
    year = {2001},
    month = {09},
    issn = {0035-8711},
    doi = {10.1046/j.1365-8711.2001.04606.x},
    url = {https://doi.org/10.1046/j.1365-8711.2001.04606.x},
    eprint = {https://academic.oup.com/mnras/article-pdf/326/1/274/3408096/326-1-274.pdf},
}

@ARTICLE{Tauris+1999,
       author = {{Tauris}, Thomas M. and {Savonije}, Gerrit J.},
        title = "{Formation of millisecond pulsars. I. Evolution of low-mass X-ray binaries with P\_orb> 2 days}",
      journal = {\aap},
         year = 1999,
        month = oct,
       volume = {350},
        pages = {928-944},
          doi = {10.48550/arXiv.astro-ph/9909147},
archivePrefix = {arXiv},
       eprint = {astro-ph/9909147},
 primaryClass = {astro-ph},
       adsurl = {https://ui.adsabs.harvard.edu/abs/1999A&A...350..928T}
}

@ARTICLE{Hui+2018,
       author = {{Hui}, C.~Y. and {Wu}, Kinwah and {Han}, Qin and {Kong}, A.~K.~H. and {Tam}, P.~H.~T.},
        title = "{On the Orbital Properties of Millisecond Pulsar Binaries}",
      journal = {\apj},
         year = 2018,
        month = sep,
       volume = {864},
       number = {1},
          eid = {30},
        pages = {30},
          doi = {10.3847/1538-4357/aad5ec},
archivePrefix = {arXiv},
       eprint = {1807.09001},
 primaryClass = {astro-ph.HE},
       adsurl = {https://ui.adsabs.harvard.edu/abs/2018ApJ...864...30H}
}

@ARTICLE{Shklovskii1970,
       author = {{Shklovskii}, I.~S.},
        title = "{Possible Causes of the Secular Increase in Pulsar Periods.}",
      journal = {\sovast},
         year = 1970,
        month = feb,
       volume = {13},
        pages = {562},
       adsurl = {https://ui.adsabs.harvard.edu/abs/1970SvA....13..562S}
}

@ARTICLE{Damour+Taylor1991,
       author = {{Damour}, Thibault and {Taylor}, J.~H.},
        title = "{On the Orbital Period Change of the Binary Pulsar PSR 1913+16}",
      journal = {\apj},
         year = 1991,
        month = jan,
       volume = {366},
        pages = {501},
          doi = {10.1086/169585},
       adsurl = {https://ui.adsabs.harvard.edu/abs/1991ApJ...366..501D}
}

@ARTICLE{Guo+2021,
       author = {{Guo}, Y.~J. and {Freire}, P.~C.~C. and {Guillemot}, L. and {Kramer}, M. and {Zhu}, W.~W. and {Wex}, N. and {McKee}, J.~W. and {Deller}, A. and {Ding}, H. and {Kaplan}, D.~L. and {Stappers}, B. and {Cognard}, I. and {Miao}, X. and {Haase}, L. and {Keith}, M. and {Ransom}, S.~M. and {Theureau}, G.},
        title = "{PSR J2222{\ensuremath{-}}0137. I. Improved physical parameters for the system}",
      journal = {\aap},
         year = 2021,
        month = oct,
       volume = {654},
          eid = {A16},
        pages = {A16},
          doi = {10.1051/0004-6361/202141450},
archivePrefix = {arXiv},
       eprint = {2107.09474},
 primaryClass = {astro-ph.HE},
       adsurl = {https://ui.adsabs.harvard.edu/abs/2021A&A...654A..16G}
}

@ARTICLE{Grav+2021,
       author = {{GRAVITY Collaboration} and {Abuter}, R. and {Amorim}, A. and {Baub{\"o}ck}, M. and {Berger}, J.~P. and {Bonnet}, H. and {Brandner}, W. and {Cl{\'e}net}, Y. and {Davies}, R. and {de Zeeuw}, P.~T. and {Dexter}, J. and {Dallilar}, Y. and {Drescher}, A. and {Eckart}, A. and {Eisenhauer}, F. and {F{\"o}rster Schreiber}, N.~M. and {Garcia}, P. and {Gao}, F. and {Gendron}, E. and {Genzel}, R. and {Gillessen}, S. and {Habibi}, M. and {Haubois}, X. and {Hei{\ss}el}, G. and {Henning}, T. and {Hippler}, S. and {Horrobin}, M. and {Jim{\'e}nez-Rosales}, A. and {Jochum}, L. and {Jocou}, L. and {Kaufer}, A. and {Kervella}, P. and {Lacour}, S. and {Lapeyr{\`e}re}, V. and {Le Bouquin}, J. -B. and {L{\'e}na}, P. and {Lutz}, D. and {Nowak}, M. and {Ott}, T. and {Paumard}, T. and {Perraut}, K. and {Perrin}, G. and {Pfuhl}, O. and {Rabien}, S. and {Rodr{\'\i}guez-Coira}, G. and {Shangguan}, J. and {Shimizu}, T. and {Scheithauer}, S. and {Stadler}, J. and {Straub}, O. and {Straubmeier}, C. and {Sturm}, E. and {Tacconi}, L.~J. and {Vincent}, F. and {von Fellenberg}, S. and {Waisberg}, I. and {Widmann}, F. and {Wieprecht}, E. and {Wiezorrek}, E. and {Woillez}, J. and {Yazici}, S. and {Young}, A. and {Zins}, G.},
        title = "{Improved GRAVITY astrometric accuracy from modeling optical aberrations}",
      journal = {\aap},
         year = 2021,
        month = mar,
       volume = {647},
          eid = {A59},
        pages = {A59},
          doi = {10.1051/0004-6361/202040208},
archivePrefix = {arXiv},
       eprint = {2101.12098},
 primaryClass = {astro-ph.GA},
       adsurl = {https://ui.adsabs.harvard.edu/abs/2021A&A...647A..59G}
}

@ARTICLE{Lazaridis+2009,
       author = {{Lazaridis}, K. and {Wex}, N. and {Jessner}, A. and {Kramer}, M. and {Stappers}, B.~W. and {Janssen}, G.~H. and {Desvignes}, G. and {Purver}, M.~B. and {Cognard}, I. and {Theureau}, G. and {Lyne}, A.~G. and {Jordan}, C.~A. and {Zensus}, J.~A.},
        title = "{Generic tests of the existence of the gravitational dipole radiation and the variation of the gravitational constant}",
      journal = {\mnras},
         year = 2009,
        month = dec,
       volume = {400},
       number = {2},
        pages = {805-814},
          doi = {10.1111/j.1365-2966.2009.15481.x},
archivePrefix = {arXiv},
       eprint = {0908.0285},
 primaryClass = {astro-ph.GA},
       adsurl = {https://ui.adsabs.harvard.edu/abs/2009MNRAS.400..805L}
}

@ARTICLE{Ocker+Cordes2026,
       author = {{Ocker}, Stella Koch and {Cordes}, James M.},
        title = "{NE2025: An Updated Electron Density Model for the Galactic Interstellar Medium}",
      journal = {\apj},
         year = 2026,
        month = may,
       volume = {1002},
       number = {1},
          eid = {3},
        pages = {3},
          doi = {10.3847/1538-4357/ae5825},
archivePrefix = {arXiv},
       eprint = {2602.11838},
 primaryClass = {astro-ph.GA},
       adsurl = {https://ui.adsabs.harvard.edu/abs/2026ApJ..1002....3O}
}

@ARTICLE{Manchester+2013,
       author = {{Manchester}, R.~N. and {Hobbs}, G. and {Bailes}, M. and {Coles}, W.~A. and {van Straten}, W. and {Keith}, M.~J. and {Shannon}, R.~M. and {Bhat}, N.~D.~R. and {Brown}, A. and {Burke-Spolaor}, S.~G. and {Champion}, D.~J. and {Chaudhary}, A. and {Edwards}, R.~T. and {Hampson}, G. and {Hotan}, A.~W. and {Jameson}, A. and {Jenet}, F.~A. and {Kesteven}, M.~J. and {Khoo}, J. and {Kocz}, J. and {Maciesiak}, K. and {Oslowski}, S. and {Ravi}, V. and {Reynolds}, J.~R. and {Sarkissian}, J.~M. and {Verbiest}, J.~P.~W. and {Wen}, Z.~L. and {Wilson}, W.~E. and {Yardley}, D. and {Yan}, W.~M. and {You}, X.~P.},
        title = "{The Parkes Pulsar Timing Array Project}",
      journal = {\pasa},
         year = 2013,
        month = jan,
       volume = {30},
          eid = {e017},
        pages = {e017},
          doi = {10.1017/pasa.2012.017},
archivePrefix = {arXiv},
       eprint = {1210.6130},
 primaryClass = {astro-ph.IM},
       adsurl = {https://ui.adsabs.harvard.edu/abs/2013PASA...30...17M}
}

@PHDTHESIS{Ransom2001,
       author = {{Ransom}, Scott Mitchell},
        title = "{New search techniques for binary pulsars}",
       school = {Harvard University, Massachusetts},
         year = 2001,
        month = jan,
       adsurl = {https://ui.adsabs.harvard.edu/abs/2001PhDT.......123R}
}

@software{Ransom2011,
       author = {{Ransom}, Scott},
        title = "{PRESTO: PulsaR Exploration and Search TOolkit}",
 howpublished = {Astrophysics Source Code Library, record ascl:1107.017},
         year = 2011,
        month = jul,
          eid = {ascl:1107.017},
archivePrefix = {ascl},
       eprint = {1107.017},
       adsurl = {https://ui.adsabs.harvard.edu/abs/2011ascl.soft07017R}
}

@ARTICLE{Ransom+2002,
       author = {{Ransom}, Scott M. and {Eikenberry}, Stephen S. and {Middleditch}, John},
        title = "{Fourier Techniques for Very Long Astrophysical Time-Series Analysis}",
      journal = {\aj},
         year = 2002,
        month = sep,
       volume = {124},
       number = {3},
        pages = {1788-1809},
          doi = {10.1086/342285},
archivePrefix = {arXiv},
       eprint = {astro-ph/0204349},
 primaryClass = {astro-ph},
       adsurl = {https://ui.adsabs.harvard.edu/abs/2002AJ....124.1788R}
}

@ARTICLE{Ridolfi+2021,
       author = {{Ridolfi}, A. and {Gautam}, T. and {Freire}, P.~C.~C. and {Ransom}, S.~M. and {Buchner}, S.~J. and {Possenti}, A. and {Venkatraman Krishnan}, V. and {Bailes}, M. and {Kramer}, M. and {Stappers}, B.~W. and {Abbate}, F. and {Barr}, E.~D. and {Burgay}, M. and {Camilo}, F. and {Corongiu}, A. and {Jameson}, A. and {Padmanabh}, P.~V. and {Vleeschower}, L. and {Champion}, D.~J. and {Chen}, W. and {Geyer}, M. and {Karastergiou}, A. and {Karuppusamy}, R. and {Parthasarathy}, A. and {Reardon}, D.~J. and {Serylak}, M. and {Shannon}, R.~M. and {Spiewak}, R.},
        title = "{Eight new millisecond pulsars from the first MeerKAT globular cluster census}",
      journal = {\mnras},
         year = 2021,
        month = jun,
       volume = {504},
       number = {1},
        pages = {1407-1426},
          doi = {10.1093/mnras/stab790},
archivePrefix = {arXiv},
       eprint = {2103.04800},
 primaryClass = {astro-ph.HE},
       adsurl = {https://ui.adsabs.harvard.edu/abs/2021MNRAS.504.1407R}
}

@ARTICLE{Wang+2024,
       author = {{Wang}, Ziteng and {Kaplan}, David L. and {Sengar}, Rahul and {Lenc}, Emil and {Zic}, Andrew and {Anumarlapudi}, Akash and {Gaensler}, B.~M. and {Hurley-Walker}, Natasha and {Murphy}, Tara and {Wang}, Yuanming},
        title = "{Discovery of a Young, Highly Scattered Pulsar PSR J1032-5804 with the Australian Square Kilometre Array Pathfinder}",
      journal = {\apj},
         year = 2024,
        month = feb,
       volume = {961},
       number = {2},
          eid = {175},
        pages = {175},
          doi = {10.3847/1538-4357/ad0fe8},
archivePrefix = {arXiv},
       eprint = {2311.14880},
 primaryClass = {astro-ph.HE},
       adsurl = {https://ui.adsabs.harvard.edu/abs/2024ApJ...961..175W}
}

@ARTICLE{Keith+2010,
       author = {{Keith}, M.~J. and {Jameson}, A. and {van Straten}, W. and {Bailes}, M. and {Johnston}, S. and {Kramer}, M. and {Possenti}, A. and {Bates}, S.~D. and {Bhat}, N.~D.~R. and {Burgay}, M. and {Burke-Spolaor}, S. and {D'Amico}, N. and {Levin}, L. and {McMahon}, Peter L. and {Milia}, S. and {Stappers}, B.~W.},
        title = "{The High Time Resolution Universe Pulsar Survey - I. System configuration and initial discoveries}",
      journal = {\mnras},
         year = 2010,
        month = dec,
       volume = {409},
       number = {2},
        pages = {619-627},
          doi = {10.1111/j.1365-2966.2010.17325.x},
archivePrefix = {arXiv},
       eprint = {1006.5744},
 primaryClass = {astro-ph.HE},
       adsurl = {https://ui.adsabs.harvard.edu/abs/2010MNRAS.409..619K}
}

@ARTICLE{Manchester+2001,
       author = {{Manchester}, R.~N. and {Lyne}, A.~G. and {Camilo}, F. and {Bell}, J.~F. and {Kaspi}, V.~M. and {D'Amico}, N. and {McKay}, N.~P.~F. and {Crawford}, F. and {Stairs}, I.~H. and {Possenti}, A. and {Kramer}, M. and {Sheppard}, D.~C.},
        title = "{The Parkes multi-beam pulsar survey - I. Observing and data analysis systems, discovery and timing of 100 pulsars}",
      journal = {\mnras},
         year = 2001,
        month = nov,
       volume = {328},
       number = {1},
        pages = {17-35},
          doi = {10.1046/j.1365-8711.2001.04751.x},
archivePrefix = {arXiv},
       eprint = {astro-ph/0106522},
 primaryClass = {astro-ph},
       adsurl = {https://ui.adsabs.harvard.edu/abs/2001MNRAS.328...17M}
}

@ARTICLE{Padmanabh+2023,
       author = {{Padmanabh}, P.~V. and {Barr}, E.~D. and {Sridhar}, S.~S. and {Rugel}, M.~R. and {Damas-Segovia}, A. and {Jacob}, A.~M. and {Balakrishnan}, V. and {Berezina}, M. and {Bernadich}, M.~C. and {Brunthaler}, A. and {Champion}, D.~J. and {Freire}, P.~C.~C. and {Khan}, S. and {Kl{\"o}ckner}, H. -R. and {Kramer}, M. and {Ma}, Y.~K. and {Mao}, S.~A. and {Men}, Y.~P. and {Menten}, K.~M. and {Sengupta}, S. and {Venkatraman Krishnan}, V. and {Wucknitz}, O. and {Wyrowski}, F. and {Bezuidenhout}, M.~C. and {Buchner}, S. and {Burgay}, M. and {Chen}, W. and {Clark}, C.~J. and {K{\"u}nkel}, L. and {Nieder}, L. and {Stappers}, B. and {Legodi}, L.~S. and {Nyamai}, M.~M.},
        title = "{The MPIfR-MeerKAT Galactic Plane Survey - I. System set-up and early results}",
      journal = {\mnras},
         year = 2023,
        month = sep,
       volume = {524},
       number = {1},
        pages = {1291-1315},
          doi = {10.1093/mnras/stad1900},
archivePrefix = {arXiv},
       eprint = {2303.09231},
 primaryClass = {astro-ph.HE},
       adsurl = {https://ui.adsabs.harvard.edu/abs/2023MNRAS.524.1291P}
}

@INPROCEEDINGS{goodman,
       author = {{Clemens}, J. Christopher and {Crain}, J. Adam and {Anderson}, Robert},
        title = "{The Goodman spectrograph}",
    booktitle = {Ground-based Instrumentation for Astronomy},
         year = 2004,
       editor = {{Moorwood}, Alan F.~M. and {Iye}, Masanori},
       series = {Society of Photo-Optical Instrumentation Engineers (SPIE) Conference Series},
       volume = {5492},
        month = sep,
        pages = {331-340},
          doi = {10.1117/12.550069},
       adsurl = {https://ui.adsabs.harvard.edu/abs/2004SPIE.5492..331C}
}

@software{astrometry,
       author = {{Lang}, Dustin and {Hogg}, David W. and {Mierle}, Keir and {Blanton}, Michael and {Roweis}, Sam},
        title = "{Astrometry.net: Astrometric calibration of images}",
 howpublished = {Astrophysics Source Code Library, record ascl:1208.001},
         year = 2012,
        month = aug,
          eid = {ascl:1208.001},
archivePrefix = {ascl},
       eprint = {1208.001},
       adsurl = {https://ui.adsabs.harvard.edu/abs/2012ascl.soft08001L}
}

@software{swarp,
       author = {{Bertin}, Emmanuel},
        title = "{SWarp: Resampling and Co-adding FITS Images Together}",
 howpublished = {Astrophysics Source Code Library, record ascl:1010.068},
         year = 2010,
        month = oct,
          eid = {ascl:1010.068},
archivePrefix = {ascl},
       eprint = {1010.068},
       adsurl = {https://ui.adsabs.harvard.edu/abs/2010ascl.soft10068B}
}

@software{sex,
       author = {{Bertin}, E. and {Arnouts}, S.},
        title = "{SExtractor: Source Extractor}",
 howpublished = {Astrophysics Source Code Library, record ascl:1010.064},
         year = 2010,
        month = oct,
          eid = {ascl:1010.064},
archivePrefix = {ascl},
       eprint = {1010.064},
       adsurl = {https://ui.adsabs.harvard.edu/abs/2010ascl.soft10064B}
}

@ARTICLE{gaiadr3,
       author = {{Gaia Collaboration} and {Vallenari}, A. and {Brown}, A.~G.~A. and {Prusti}, T. and {de Bruijne}, J.~H.~J. and {Arenou}, F. and {Babusiaux}, C. and {Biermann}, M. and {Creevey}, O.~L. and {Ducourant}, C. and {Evans}, D.~W. and {Eyer}, L. and {Guerra}, R. and {Hutton}, A. and {Jordi}, C. and {Klioner}, S.~A. and {Lammers}, U.~L. and {Lindegren}, L. and {Luri}, X. and {Mignard}, F. and {Panem}, C. and {Pourbaix}, D. and {Randich}, S. and {Sartoretti}, P. and {Soubiran}, C. and {Tanga}, P. and {Walton}, N.~A. and {Bailer-Jones}, C.~A.~L. and {Bastian}, U. and {Drimmel}, R. and {Jansen}, F. and {Katz}, D. and {Lattanzi}, M.~G. and {van Leeuwen}, F. and {Bakker}, J. and {Cacciari}, C. and {Casta{\~n}eda}, J. and {De Angeli}, F. and {Fabricius}, C. and {Fouesneau}, M. and {Fr{\'e}mat}, Y. and {Galluccio}, L. and {Guerrier}, A. and {Heiter}, U. and {Masana}, E. and {Messineo}, R. and {Mowlavi}, N. and {Nicolas}, C. and {Nienartowicz}, K. and {Pailler}, F. and {Panuzzo}, P. and {Riclet}, F. and {Roux}, W. and {Seabroke}, G.~M. and {Sordo}, R. and {Th{\'e}venin}, F. and {Gracia-Abril}, G. and {Portell}, J. and {Teyssier}, D. and {Altmann}, M. and {Andrae}, R. and {Audard}, M. and {Bellas-Velidis}, I. and {Benson}, K. and {Berthier}, J. and {Blomme}, R. and {Burgess}, P.~W. and {Busonero}, D. and {Busso}, G. and {C{\'a}novas}, H. and {Carry}, B. and {Cellino}, A. and {Cheek}, N. and {Clementini}, G. and {Damerdji}, Y. and {Davidson}, M. and {de Teodoro}, P. and {Nu{\~n}ez Campos}, M. and {Delchambre}, L. and {Dell'Oro}, A. and {Esquej}, P. and {Fern{\'a}ndez-Hern{\'a}ndez}, J. and {Fraile}, E. and {Garabato}, D. and {Garc{\'\i}a-Lario}, P. and {Gosset}, E. and {Haigron}, R. and {Halbwachs}, J.-L. and {Hambly}, N.~C. and {Harrison}, D.~L. and {Hern{\'a}ndez}, J. and {Hestroffer}, D. and {Hodgkin}, S.~T. and {Holl}, B. and {Jan{\ss}en}, K. and {Jevardat de Fombelle}, G. and {Jordan}, S. and {Krone-Martins}, A. and {Lanzafame}, A.~C. and {L{\"o}ffler}, W. and {Marchal}, O. and {Marrese}, P.~M. and {Moitinho}, A. and {Muinonen}, K. and {Osborne}, P. and {Pancino}, E. and {Pauwels}, T. and {Recio-Blanco}, A. and {Reyl{\'e}}, C. and {Riello}, M. and {Rimoldini}, L. and {Roegiers}, T. and {Rybizki}, J. and {Sarro}, L.~M. and {Siopis}, C. and {Smith}, M. and {Sozzetti}, A. and {Utrilla}, E. and {van Leeuwen}, M. and {Abbas}, U. and {{\'A}brah{\'a}m}, P. and {Abreu Aramburu}, A. and {Aerts}, C. and {Aguado}, J.~J. and {Ajaj}, M. and {Aldea-Montero}, F. and {Altavilla}, G. and {{\'A}lvarez}, M.~A. and {Alves}, J. and {Anders}, F. and {Anderson}, R.~I. and {Anglada Varela}, E. and {Antoja}, T. and {Baines}, D. and {Baker}, S.~G. and {Balaguer-N{\'u}{\~n}ez}, L. and {Balbinot}, E. and {Balog}, Z. and {Barache}, C. and {Barbato}, D. and {Barros}, M. and {Barstow}, M.~A. and {Bartolom{\'e}}, S. and {Bassilana}, J.-L. and {Bauchet}, N. and {Becciani}, U. and {Bellazzini}, M. and {Berihuete}, A. and {Bernet}, M. and {Bertone}, S. and {Bianchi}, L. and {Binnenfeld}, A. and {Blanco-Cuaresma}, S. and {Blazere}, A. and {Boch}, T. and {Bombrun}, A. and {Bossini}, D. and {Bouquillon}, S. and {Bragaglia}, A. and {Bramante}, L. and {Breedt}, E. and {Bressan}, A. and {Brouillet}, N. and {Brugaletta}, E. and {Bucciarelli}, B. and {Burlacu}, A. and {Butkevich}, A.~G. and {Buzzi}, R. and {Caffau}, E. and {Cancelliere}, R. and {Cantat-Gaudin}, T. and {Carballo}, R. and {Carlucci}, T. and {Carnerero}, M.~I. and {Carrasco}, J.~M. and {Casamiquela}, L. and {Castellani}, M. and {Castro-Ginard}, A. and {Chaoul}, L. and {Charlot}, P. and {Chemin}, L. and {Chiaramida}, V. and {Chiavassa}, A. and {Chornay}, N. and {Comoretto}, G. and {Contursi}, G. and {Cooper}, W.~J. and {Cornez}, T. and {Cowell}, S. and {Crifo}, F. and {Cropper}, M. and {Crosta}, M. and {Crowley}, C. and {Dafonte}, C. and {Dapergolas}, A. and {David}, M. and {David}, P. and {de Laverny}, P. and {De Luise}, F. and {De March}, R.},
        title = "{Gaia Data Release 3. Summary of the content and survey properties}",
      journal = {\aap},
         year = 2023,
        month = jun,
       volume = {674},
          eid = {A1},
        pages = {A1},
          doi = {10.1051/0004-6361/202243940},
archivePrefix = {arXiv},
       eprint = {2208.00211},
 primaryClass = {astro-ph.GA},
       adsurl = {https://ui.adsabs.harvard.edu/abs/2023A&A...674A...1G}
}

@ARTICLE{1998ApJ...500..525S,
       author = {{Schlegel}, David J. and {Finkbeiner}, Douglas P. and {Davis}, Marc},
        title = "{Maps of Dust Infrared Emission for Use in Estimation of Reddening and Cosmic Microwave Background Radiation Foregrounds}",
      journal = {\apj},
         year = 1998,
        month = jun,
       volume = {500},
       number = {2},
        pages = {525-553},
          doi = {10.1086/305772},
archivePrefix = {arXiv},
       eprint = {astro-ph/9710327},
 primaryClass = {astro-ph},
       adsurl = {https://ui.adsabs.harvard.edu/abs/1998ApJ...500..525S}
}

@article{Green2018, doi = {10.21105/joss.00695}, url = {https://doi.org/10.21105/joss.00695}, year = {2018}, publisher = {The Open Journal}, volume = {3}, number = {26}, pages = {695}, author = {Green, Gregory M.}, title = {dustmaps: A Python interface for maps of interstellar dust}, journal = {Journal of Open Source Software} }

@ARTICLE{2006AJ....132.1221H,
       author = {{Holberg}, J.~B. and {Bergeron}, Pierre},
        title = "{Calibration of Synthetic Photometry Using DA White Dwarfs}",
      journal = {\aj},
         year = 2006,
        month = sep,
       volume = {132},
       number = {3},
        pages = {1221-1233},
          doi = {10.1086/505938},
       adsurl = {https://ui.adsabs.harvard.edu/abs/2006AJ....132.1221H}
}

@ARTICLE{1995PASP..107.1047B,
       author = {{Bergeron}, P. and {Wesemael}, F. and {Beauchamp}, A.},
        title = "{Photometric Calibration of Hydrogen- and Helium-Rich White Dwarf Models}",
      journal = {\pasp},
         year = 1995,
        month = nov,
       volume = {107},
        pages = {1047},
          doi = {10.1086/133661},
       adsurl = {https://ui.adsabs.harvard.edu/abs/1995PASP..107.1047B}
}

@article{He_2013,
doi = {10.1088/0004-637X/768/1/64},
url = {https://doi.org/10.1088/0004-637X/768/1/64},
year = {2013},
month = {apr},
publisher = {The American Astronomical Society},
volume = {768},
number = {1},
pages = {64},
author = {He, C. and Ng, C.-Y. and Kaspi, V. M.},
title = {THE CORRELATION BETWEEN DISPERSION MEASURE AND X-RAY COLUMN DENSITY FROM RADIO PULSARS},
journal = {The Astrophysical Journal}
}

@ARTICLE{1995A&A...293..889P,
       author = {{Predehl}, P. and {Schmitt}, J.~H.~M.~M.},
        title = "{X-raying the interstellar medium: ROSAT observations of dust scattering halos.}",
      journal = {\aap},
         year = 1995,
        month = jan,
       volume = {293},
        pages = {889-905},
       adsurl = {https://ui.adsabs.harvard.edu/abs/1995A&A...293..889P}
}

@ARTICLE{NANOGrav+2025,
       author = {{Agazie}, Gabriella and {Anumarlapudi}, Akash and {Archibald}, Anne M. and {Arzoumanian}, Zaven and {Baier}, Jeremy George and {Baker}, Paul T. and {B{\'e}csy}, Bence and {Blecha}, Laura and {Brazier}, Adam and {Brook}, Paul R. and {Burke-Spolaor}, Sarah and {Casey-Clyde}, J. Andrew and {Charisi}, Maria and {Chatterjee}, Shami and {Chatziioannou}, Katerina and {Cohen}, Tyler and {Cordes}, James M. and {Cornish}, Neil J. and {Crawford}, Fronefield and {Cromartie}, H. Thankful and {Crowter}, Kathryn and {Decesar}, Megan E. and {Demorest}, Paul B. and {Deng}, Heling and {Dey}, Lankeswar and {Dolch}, Timothy and {Ferrara}, Elizabeth C. and {Fiore}, William and {Fonseca}, Emmanuel and {Freedman}, Gabriel E. and {Gardiner}, Emiko C. and {Garver-Daniels}, Nate and {Gentile}, Peter A. and {Gersbach}, Kyle A. and {Glaser}, Joseph and {Good}, Deborah C. and {G{\"u}ltekin}, Kayhan and {Hazboun}, Jeffrey S. and {Jennings}, Ross J. and {Johnson}, Aaron D. and {Jones}, Megan L. and {Kaiser}, Andrew R. and {Kaplan}, David L. and {Kelley}, Luke Zoltan and {Kerr}, Matthew and {Key}, Joey S. and {Laal}, Nima and {Lam}, Michael T. and {Lamb}, William G. and {Larsen}, Bjorn and {Joseph}, T. and {Lazio}, W. and {Lewandowska}, Natalia and {Liu}, Tingting and {Lorimer}, Duncan R. and {Luo}, Jing and {Lynch}, Ryan S. and {Ma}, Chung-Pei and {Madison}, Dustin R. and {McEwen}, Alexander and {McKee}, James W. and {McLaughlin}, Maura A. and {McMann}, Natasha and {Meyers}, Bradley W. and {Meyers}, Patrick M. and {Mingarelli}, Chiara M.~F. and {Mitridate}, Andrea and {Ng}, Cherry and {Nice}, David J. and {Ocker}, Stella Koch and {Olum}, Ken D. and {Pennucci}, Timothy T. and {Perera}, Benetge B.~P. and {Pol}, Nihan S. and {Radovan}, Henri A. and {Ransom}, Scott M. and {Ray}, Paul S. and {Romano}, Joseph D. and {Runnoe}, Jessie C. and {Saffer}, Alexander and {Sardesai}, Shashwat C. and {Schmiedekamp}, Ann and {Schmiedekamp}, Carl and {Schmitz}, Kai and {Shapiro-Albert}, Brent J. and {Siemens}, Xavier and {Simon}, Joseph and {Siwek}, Magdalena S. and {Fiscella}, Sophia V. Sosa and {Stairs}, Ingrid H. and {Stinebring}, Daniel R. and {Stovall}, Kevin and {Susobhanan}, Abhimanyu and {Swiggum}, Joseph K. and {Taylor}, Stephen R. and {Turner}, Jacob E. and {Unal}, Caner and {Vallisneri}, Michele and {Vigeland}, Sarah J. and {Wahl}, Haley M. and {Witt}, Caitlin A. and {Wright}, David and {Young}, Olivia and {Nanograv Collaboration}},
        title = "{The NANOGrav 15 yr dataset: Posterior predictive checks for gravitational-wave detection with pulsar timing arrays}",
      journal = {\prd},
         year = 2025,
        month = feb,
       volume = {111},
       number = {4},
          eid = {042011},
        pages = {042011},
          doi = {10.1103/PhysRevD.111.042011},
archivePrefix = {arXiv},
       eprint = {2407.20510},
 primaryClass = {astro-ph.HE},
       adsurl = {https://ui.adsabs.harvard.edu/abs/2025PhRvD.111d2011A}
}

@ARTICLE{EPTA+2023,
       author = {{EPTA Collaboration} and {InPTA Collaboration} and {Antoniadis}, J. and {Arumugam}, P. and {Arumugam}, S. and {Babak}, S. and {Bagchi}, M. and {Bak Nielsen}, A.-S. and {Bassa}, C.~G. and {Bathula}, A. and {Berthereau}, A. and {Bonetti}, M. and {Bortolas}, E. and {Brook}, P.~R. and {Burgay}, M. and {Caballero}, R.~N. and {Chalumeau}, A. and {Champion}, D.~J. and {Chanlaridis}, S. and {Chen}, S. and {Cognard}, I. and {Dandapat}, S. and {Deb}, D. and {Desai}, S. and {Desvignes}, G. and {Dhanda-Batra}, N. and {Dwivedi}, C. and {Falxa}, M. and {Ferdman}, R.~D. and {Franchini}, A. and {Gair}, J.~R. and {Goncharov}, B. and {Gopakumar}, A. and {Graikou}, E. and {Grie{\ss}meier}, J.-M. and {Guillemot}, L. and {Guo}, Y.~J. and {Gupta}, Y. and {Hisano}, S. and {Hu}, H. and {Iraci}, F. and {Izquierdo-Villalba}, D. and {Jang}, J. and {Jawor}, J. and {Janssen}, G.~H. and {Jessner}, A. and {Joshi}, B.~C. and {Kareem}, F. and {Karuppusamy}, R. and {Keane}, E.~F. and {Keith}, M.~J. and {Kharbanda}, D. and {Kikunaga}, T. and {Kolhe}, N. and {Kramer}, M. and {Krishnakumar}, M.~A. and {Lackeos}, K. and {Lee}, K.~J. and {Liu}, K. and {Liu}, Y. and {Lyne}, A.~G. and {McKee}, J.~W. and {Maan}, Y. and {Main}, R.~A. and {Mickaliger}, M.~B. and {Ni{\c{t}}u}, I.~C. and {Nobleson}, K. and {Paladi}, A.~K. and {Parthasarathy}, A. and {Perera}, B.~B.~P. and {Perrodin}, D. and {Petiteau}, A. and {Porayko}, N.~K. and {Possenti}, A. and {Prabu}, T. and {Quelquejay Leclere}, H. and {Rana}, P. and {Samajdar}, A. and {Sanidas}, S.~A. and {Sesana}, A. and {Shaifullah}, G. and {Singha}, J. and {Speri}, L. and {Spiewak}, R. and {Srivastava}, A. and {Stappers}, B.~W. and {Surnis}, M. and {Susarla}, S.~C. and {Susobhanan}, A. and {Takahashi}, K. and {Tarafdar}, P. and {Theureau}, G. and {Tiburzi}, C. and {van der Wateren}, E. and {Vecchio}, A. and {Venkatraman Krishnan}, V. and {Verbiest}, J.~P.~W. and {Wang}, J. and {Wang}, L. and {Wu}, Z.},
        title = "{The second data release from the European Pulsar Timing Array. III. Search for gravitational wave signals}",
      journal = {\aap},
         year = 2023,
        month = oct,
       volume = {678},
          eid = {A50},
        pages = {A50},
          doi = {10.1051/0004-6361/202346844},
archivePrefix = {arXiv},
       eprint = {2306.16214},
 primaryClass = {astro-ph.HE},
       adsurl = {https://ui.adsabs.harvard.edu/abs/2023A&A...678A..50E}
}

@ARTICLE{CPTA+2023,
       author = {{Xu}, Heng and {Chen}, Siyuan and {Guo}, Yanjun and {Jiang}, Jinchen and {Wang}, Bojun and {Xu}, Jiangwei and {Xue}, Zihan and {Caballero}, R. Nicolas and {Yuan}, Jianping and {Xu}, Yonghua and {Wang}, Jingbo and {Hao}, Longfei and {Luo}, Jingtao and {Lee}, Kejia and {Han}, Jinlin and {Jiang}, Peng and {Shen}, Zhiqiang and {Wang}, Min and {Wang}, Na and {Xu}, Renxin and {Wu}, Xiangping and {Manchester}, Richard and {Qian}, Lei and {Guan}, Xin and {Huang}, Menglin and {Sun}, Chun and {Zhu}, Yan},
        title = "{Searching for the Nano-Hertz Stochastic Gravitational Wave Background with the Chinese Pulsar Timing Array Data Release I}",
      journal = {Research in Astronomy and Astrophysics},
         year = 2023,
        month = jul,
       volume = {23},
       number = {7},
          eid = {075024},
        pages = {075024},
          doi = {10.1088/1674-4527/acdfa5},
archivePrefix = {arXiv},
       eprint = {2306.16216},
 primaryClass = {astro-ph.HE},
       adsurl = {https://ui.adsabs.harvard.edu/abs/2023RAA....23g5024X}
}

@ARTICLE{Agazie+2023SMBHBs,
       author = {{Agazie}, Gabriella and {Anumarlapudi}, Akash and {Archibald}, Anne M. and {Arzoumanian}, Zaven and {Baker}, Paul T. and {B{\'e}csy}, Bence and {Blecha}, Laura and {Brazier}, Adam and {Brook}, Paul R. and {Burke-Spolaor}, Sarah and {Case}, Robin and {Casey-Clyde}, J. Andrew and {Charisi}, Maria and {Chatterjee}, Shami and {Cohen}, Tyler and {Cordes}, James M. and {Cornish}, Neil J. and {Crawford}, Fronefield and {Cromartie}, H. Thankful and {Crowter}, Kathryn and {Decesar}, Megan E. and {Demorest}, Paul B. and {Digman}, Matthew C. and {Dolch}, Timothy and {Drachler}, Brendan and {Ferrara}, Elizabeth C. and {Fiore}, William and {Fonseca}, Emmanuel and {Freedman}, Gabriel E. and {Garver-Daniels}, Nate and {Gentile}, Peter A. and {Glaser}, Joseph and {Good}, Deborah C. and {G{\"u}ltekin}, Kayhan and {Hazboun}, Jeffrey S. and {Hourihane}, Sophie and {Jennings}, Ross J. and {Johnson}, Aaron D. and {Jones}, Megan L. and {Kaiser}, Andrew R. and {Kaplan}, David L. and {Kelley}, Luke Zoltan and {Kerr}, Matthew and {Key}, Joey S. and {Laal}, Nima and {Lam}, Michael T. and {Lamb}, William G. and {Lazio}, T. Joseph W. and {Lewandowska}, Natalia and {Liu}, Tingting and {Lorimer}, Duncan R. and {Luo}, Jing and {Lynch}, Ryan S. and {Ma}, Chung-Pei and {Madison}, Dustin R. and {McEwen}, Alexander and {McKee}, James W. and {McLaughlin}, Maura A. and {McMann}, Natasha and {Meyers}, Bradley W. and {Meyers}, Patrick M. and {Mingarelli}, Chiara M.~F. and {Mitridate}, Andrea and {Ng}, Cherry and {Nice}, David J. and {Ocker}, Stella Koch and {Olum}, Ken D. and {Pennucci}, Timothy T. and {Perera}, Benetge B.~P. and {Petrov}, Polina and {Pol}, Nihan S. and {Radovan}, Henri A. and {Ransom}, Scott M. and {Ray}, Paul S. and {Romano}, Joseph D. and {Sardesai}, Shashwat C. and {Schmiedekamp}, Ann and {Schmiedekamp}, Carl and {Schmitz}, Kai and {Shapiro-Albert}, Brent J. and {Siemens}, Xavier and {Simon}, Joseph and {Siwek}, Magdalena S. and {Stairs}, Ingrid H. and {Stinebring}, Daniel R. and {Stovall}, Kevin and {Susobhanan}, Abhimanyu and {Swiggum}, Joseph K. and {Taylor}, Jacob and {Taylor}, Stephen R. and {Turner}, Jacob E. and {Unal}, Caner and {Vallisneri}, Michele and {van Haasteren}, Rutger and {Vigeland}, Sarah J. and {Wahl}, Haley M. and {Witt}, Caitlin A. and {Young}, Olivia and {Nanograv Collaboration}},
        title = "{The NANOGrav 15 yr Data Set: Bayesian Limits on Gravitational Waves from Individual Supermassive Black Hole Binaries}",
      journal = {\apjl},
         year = 2023,
        month = jul,
       volume = {951},
       number = {2},
          eid = {L50},
        pages = {L50},
          doi = {10.3847/2041-8213/ace18a},
archivePrefix = {arXiv},
       eprint = {2306.16222},
 primaryClass = {astro-ph.HE},
       adsurl = {https://ui.adsabs.harvard.edu/abs/2023ApJ...951L..50A}
}

@INPROCEEDINGS{Luo+2018,
       author = {{Luo}, Jing and {Ransom}, Scott M. and {Demorest}, Paul and {Ray}, Paul S. and {Stovall}, Kevin and {Jenet}, Fredrick and {Ellis}, Justin and {van Haasteren}, Rutger and {Bachetti}, Matteo and {NANOGrav PINT developer Team}},
        title = "{PINT, A Modern Software Package for Pulsar Timing}",
    booktitle = {American Astronomical Society Meeting Abstracts \#231},
         year = 2018,
       series = {American Astronomical Society Meeting Abstracts},
       volume = {231},
        month = jan,
          eid = {453.09},
        pages = {453.09},
       adsurl = {https://ui.adsabs.harvard.edu/abs/2018AAS...23145309L}
}

@ARTICLE{Kerr+2012,
       author = {{Kerr}, M. and {Camilo}, F. and {Johnson}, T.~J. and {Ferrara}, E.~C. and {Guillemot}, L. and {Harding}, A.~K. and {Hessels}, J. and {Johnston}, S. and {Keith}, M. and {Kramer}, M. and {Ransom}, S.~M. and {Ray}, P.~S. and {Reynolds}, J.~E. and {Sarkissian}, J. and {Wood}, K.~S.},
        title = "{Five New Millisecond Pulsars from a Radio Survey of 14 Unidentified Fermi-LAT Gamma-Ray Sources}",
      journal = {\apjl},
         year = 2012,
        month = mar,
       volume = {748},
       number = {1},
          eid = {L2},
        pages = {L2},
          doi = {10.1088/2041-8205/748/1/L2},
archivePrefix = {arXiv},
       eprint = {1201.5160},
 primaryClass = {astro-ph.HE},
       adsurl = {https://ui.adsabs.harvard.edu/abs/2012ApJ...748L...2K}
}

@ARTICLE{Cromartie+2016,
       author = {{Cromartie}, H.~T. and {Camilo}, F. and {Kerr}, M. and {Deneva}, J.~S. and {Ransom}, S.~M. and {Ray}, P.~S. and {Ferrara}, E.~C. and {Michelson}, P.~F. and {Wood}, K.~S.},
        title = "{Six New Millisecond Pulsars from Arecibo Searches of Fermi Gamma-Ray Sources}",
      journal = {\apj},
         year = 2016,
        month = mar,
       volume = {819},
       number = {1},
          eid = {34},
        pages = {34},
          doi = {10.3847/0004-637X/819/1/34},
archivePrefix = {arXiv},
       eprint = {1601.05343},
 primaryClass = {astro-ph.HE},
       adsurl = {https://ui.adsabs.harvard.edu/abs/2016ApJ...819...34C}
}

@ARTICLE{Bangale+2024,
       author = {{Bangale}, P. and {Bhattacharyya}, B. and {Camilo}, F. and {Clark}, C.~J. and {Cognard}, I. and {DeCesar}, M.~E. and {Ferrara}, E.~C. and {Gentile}, P. and {Guillemot}, L. and {Hessels}, J.~W.~T. and {Johnson}, T.~J. and {Kerr}, M. and {McLaughlin}, M.~A. and {Nieder}, L. and {Ransom}, S.~M. and {Ray}, P.~S. and {Roberts}, M.~S.~E. and {Roy}, J. and {Sanpa-arsa}, S. and {Theureau}, G. and {Wolff}, M.~T.},
        title = "{A 350 MHz Green Bank Telescope Survey of Unassociated Fermi LAT Sources: Discovery and Timing of 10 Millisecond Pulsars}",
      journal = {\apj},
         year = 2024,
        month = may,
       volume = {966},
       number = {2},
          eid = {161},
        pages = {161},
          doi = {10.3847/1538-4357/ad2994},
archivePrefix = {arXiv},
       eprint = {2402.09366},
 primaryClass = {astro-ph.HE},
       adsurl = {https://ui.adsabs.harvard.edu/abs/2024ApJ...966..161B}
}
\bibliographystyle{aasjournal}

\end{document}